\documentclass[preprint,
superscriptaddress,
nobibnotes,
amsmath,amssymb,
aps,
prx,onecolumn
]{revtex4-2}

\usepackage{caption}

\usepackage{graphicx}
\usepackage{dcolumn}
\usepackage{bm}
\usepackage{graphicx,epsfig,epstopdf}
\usepackage{amsmath} 
\usepackage{bm}
\usepackage{color}
\usepackage{subfig}
\usepackage{amssymb} 
\DeclareMathAlphabet{\mathcal}{OMS}{cmsy}{m}{n}
\newcommand{\be}{\begin{eqnarray}}
\newcommand{\ee}{\end{eqnarray}}
\newcommand{\bc}{\begin{center}}
\newcommand{\ec}{\end{center}}

\newcommand{\mrm}{\mathrm}
\newcommand{\mbf}{\mathbf}
\newcommand{\nn}{\nonumber}

\newcommand{\ex}[1]{\langle#1\rangle}

\usepackage{centernot}
\usepackage{cancel}

\begin{document}

\title{Theory of high-energy correlated multiphoton x-ray diffraction for synchrotron-radiation sources}

\author{Arunangshu Debnath}
\email{arunangshu.debnath@desy.de}
\affiliation{Center for Free-Electron Laser Science CFEL, Deutsches Elektronen-Synchrotron DESY, Notkestr. 85, 22607 Hamburg, Germany}

\author{Robin Santra}
\email{robin.santra@desy.de}
\affiliation{Center for Free-Electron Laser Science CFEL, Deutsches Elektronen-Synchrotron DESY, Notkestr. 85, 22607 Hamburg, Germany}
\affiliation{Department of Physics, Universit\"{a}t Hamburg, Notkestr. 9-11, 22607 Hamburg, Germany}


\begin{abstract}
We present a theoretical formulation for the multiphoton diffraction phenomenology in the nonrelativistic limit, suitable for interpreting high-energy x-ray diffraction measurements using synchrotron radiation sources. A hierarchy of approximations and the systematic analysis of limiting cases are presented. A convolutional representation of the diffraction signal allows classification of the physical resources contributing to the correlation signatures.
The formulation is intended for developing a theoretical description capable of describing plausible absence or presence of correlation signatures in elastic and inelastic diffractive scattering. Interpreting these correlation signatures in terms of the incoming field modulated many-body electronic density correlations provides a unique perspective for structural imaging studies. More essentially, it offers a framework necessary for theoretical developments of associated reconstruction algorithms.
\end{abstract}

\maketitle

\section{Introduction}
In the x-ray photon-induced diffractive scattering measurements where signals are monitored in the far-field zone, one of the central goals lies in extracting information about the hitherto unknown structure of the materials. In these measurements, high-resolution structural information in the real space are routinely obtained by monitoring the photon scattering patterns and applying suitable inversion schemes \cite{chapman2006femtosecond, gaffney2007imaging, young2018roadmap, marchesini2003coherent}. Obtaining atomically resolved structural information, which has been one of the central endeavors of the x-ray crystallographic measurements, has seen enormous progress in recent years. Such improvements in precision and resolution were facilitated by the deployment of advanced light sources \cite{kotani2001resonant, yoneda2015atomic, comin2016resonant, rouxel2021signatures, dixit2012imaging, feldhaus2005x, mukamel2013multidimensional, cornacchia2004future, bennett2014time}.
For structural studies, hard x-ray sources whose wavelength corresponds to the atomic resolution are preferred as a tool. The fact that the probe wavelength is comparable to the typical length scales related to relative distances between the scattering centers (e.g., $1 \text{\AA}$ corresponds to $12.4 \,\text{keV}$) allows imaging at the atomic resolution. The operational regime of the hard x-ray sources corresponds to the photon energies ranging from several keV up to hundreds of keV \cite{als2011elements}.
The higher brilliance and the coherence of the present-day synchrotron radiation sources available at storage rings or free-electron lasers (FEL) allow fewer-shot measurements for non-reproducible objects. Storage rings, owing to their peak brightness being several orders of magnitude lower than the FEL sources, typically operate at lower flux conditions. Therefore, even though a typical measurement scheme may involve multiple pulses, the mean number of photons deployed in each shot is comparatively lower. As a consequence, the propensity of scattering events leading to ionization during individual pulse interaction, per atom remains lower compared to FEL sources. The higher brilliance of the latter, which aids the signal observation, also induces scattering processes that involve multiple, correlated photon-matter interactions. In this paper, we focus on the multiphoton nature of the diffractive scattering at the storage-ring based synchrotron radiation sources \cite{helliwell1984synchrotron, freund1989x}.
In relevant situations, the full statistical distribution of the scattered photons involving all the higher-order moments is required for extracting information about the full set of diffractive scattering events.
The detection schemes based on single-photon monitoring, which effectively observes the reduced many-particle density operator of the scattered photon field, may not be an adequate quantity. It calls for a theoretical formulation that can systematically account for a multiphoton scattering and detection scenario and, more importantly, examines the physical situations which may enable the validity of a truncated subspace monitoring of the signal. Investigating the cases where a particular reduced space detection may hold also offers rationale in terms of the underlying structural features of specific materials.
In typical measurements, pixel-resolved diffracted photon intensities are sampled in the reciprocal space of the spatial variables. Following several acquisition steps, these outcomes are composed together to generate diffraction patterns which, in turn, correspond to the parametric variations of the electronic properties of scattering centers in real space. 
The signal is routinely interpreted in terms of the time-dependent, many-body electronic density of the scattering centers \cite{cao1998ultrafast, ben1997ultrafast, dashti2020retrieving}.
Consequently, while inverting the diffraction pattern, the commonly deployed inversion schemes assume that the diffraction patterns have originated solely from the time-dependent electronic density snapshots.
Previous studies have demonstrated a generalization of this phenomenology by establishing that in the one-photon diffractive scattering measurements, the signal is in fact proportional to the many-body electronic density-density correlation function \cite{dixit2012imaging, dixit2017time, coridan2012dynamics, asban2019quantum}. In this communication, we establish a further generalization by extending it to the case of $n$-photon diffractive scattering scenario.\\
In Sec. II, we introduce the Hamiltonian and develop the integral equation-based formulation for multiphoton diffractive scattering. We present a detailed analysis for the interpretation of such signals involving the higher-order correlation functions of the incoming sources, electronic density, and detection modes.
Subsequently, in Sec. III, we analyze various plausible physical features responsible for correlation signatures in the signal. We systematically develop a hierarchy of approximations, discuss their validity, and recover, in the asymptotic limits, the commonly used expressions. A set of physical conditions that may justify the usage of such a limit are described. It will be shown that the factorization of correlation functions, often applicable, may not be a generic prescription for interpreting the observed signal.
The paper concludes with a perspective and outlook on the theory, a summary of the assumptions, presented in Sec. IV.
\begin{figure}[ht]
\centering
\includegraphics[width=.48\textwidth]{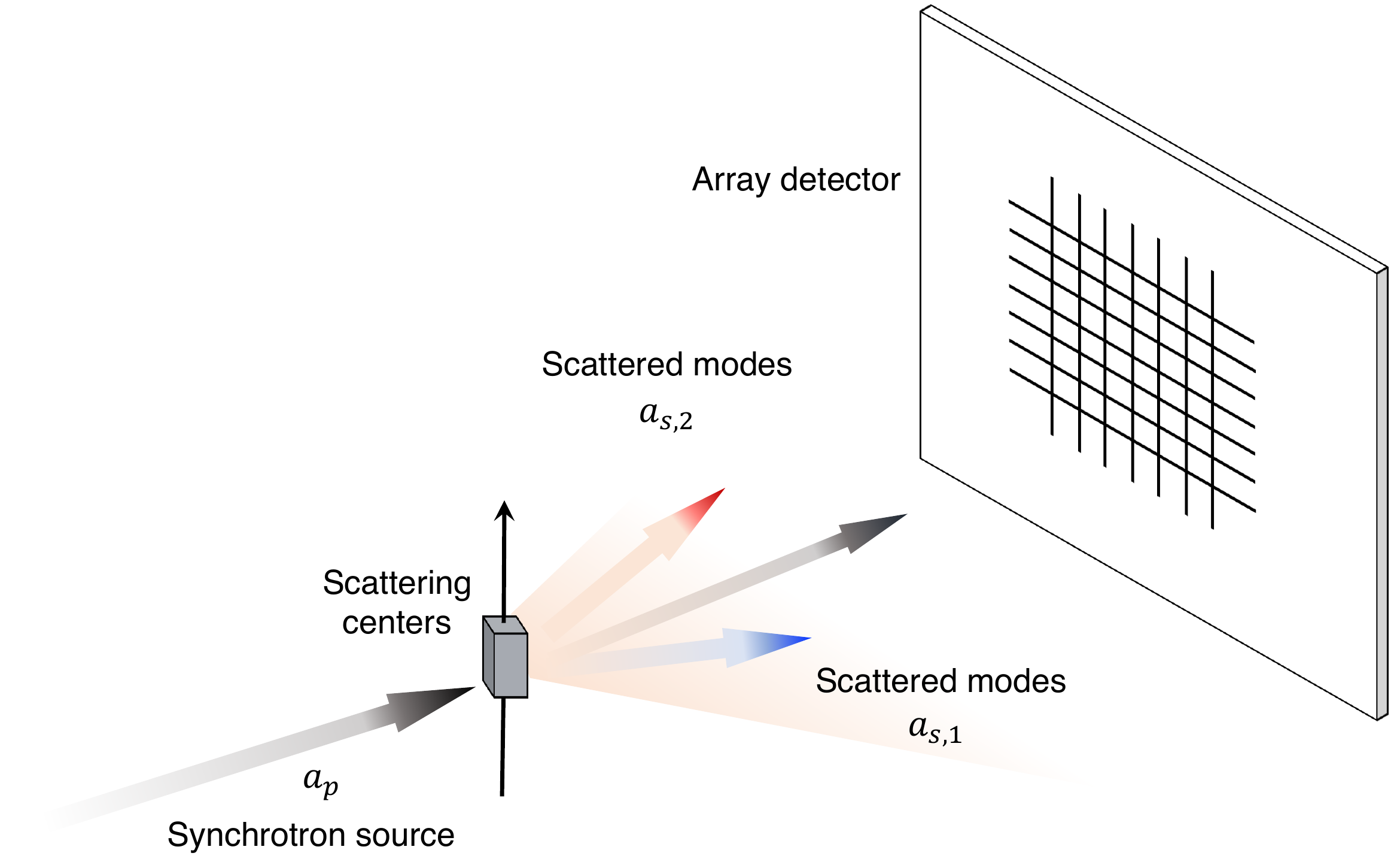}
\caption{A schematic diagram describing typical multiphoton diffraction events with array detection, which is considered in the paper. The scattered modes contain a wealth of information regarding correlations among the incoming field-modulated scattering centers. These can be revealed by focusing on the multiphoton diffraction characteristics.
} \label{fig:diffdiag}
\end{figure}
\section{Multiphoton diffraction phenomenology}
\subsection{Hamiltonian}
Multiphoton diffraction phenomenology involving interaction between electronic matter and high-energy x-ray photons in the nonrelativistic limit is described by the minimal-coupling Hamiltonian $H_{} = H_{\mathrm{matter}}+ H_{\mathrm{int}}+ H_{\mathrm{field}}$ \cite{craig1998molecular, fetter1971quantum, santra2008concepts}. The Hamiltonian components are presented below using atomic units,
\begin{align}
    & H_{\mathrm{matter}} = \sum_{\sigma}^{} \int d\mbf{r}  \psi_{}^{\dag}(\sigma \mbf{r} )\Big(-\nabla^2/2 - \sum_n \frac{ Z_n}{|\mbf{r}-\mbf{R}_n|}\Big)\psi_{}^{}(\sigma \mbf{r} )
    \nn\\
    &+\sum_{\sigma \sigma' }^{}\int \int d\mbf{r}  d\mbf{r} '  \psi_{}^\dag(\sigma \mbf{r} )\psi_{}^\dag(\sigma' \mbf{r} ')(1/2) \frac{1}{|\mbf{r}-\mbf{r}'|}\psi_{}^{}(\sigma' \mbf{r} ')\psi_{}^{}(\sigma \mbf{r}),\nn\\
    &H_{\mathrm{int}} =\sum_{\sigma}^{} \alpha\int d \mbf{r}  \psi_{}^\dagger(\sigma \mbf{r} )\mbf{A}(\mbf{r})\cdot(-i \nabla)  \psi_{}^{}(\sigma \mbf{r} ) \nn\\
    &+\sum_{\sigma}^{}\frac{\alpha^2}{2}  \int d\mbf{r}  \psi_{}^\dagger(\sigma \mbf{r} ) \mbf{A}^2(\mbf{r} ) \psi_{}^{}(\sigma \mbf{r} ) ,\nonumber\\
    & H_{\mathrm{field}} =    \sum_{\mbf{k},\mu}  \omega_{\mbf{k}}^{} (a_{\mbf{k},\mu}^{\dag} a_{\mbf{k},\mu}^{}+1/2) .
\end{align}\label{eqn:ham}
In the above, the first two terms in the $H_{\mathrm{matter}}$ correspond to the electronic matter modes. The first term is the one-body operator consisting of electron kinetic energy and the electron-nuclear potential, whereas the second term is a two-body operator describing the electron-electron interactions.
We introduced the electronic field operators $\psi_{}^{}(\sigma_{} \mbf{r} )$ (associated spin index is denoted by $\sigma$) which follow fermionic anti-commutation relations $\{\psi_{}^{}(\sigma_{} \mbf{r} ),\psi_{}^{\dag}(\sigma_{}' \mbf{r}' )\}_+ =\delta_{\sigma\sigma'} \delta_{}(\mbf{r}-\mbf{r}')$.
The external field-matter interaction Hamiltonian $H_{\mathrm{int}}$ is classified into two terms, namely $H_{\mathrm{int}}^{(1)}$ and $ H_{\mathrm{int}}^{(2)}$, given in the third and fourth line, respectively. The term $H_{\mathrm{int}}^{(1)}$ containing $\mbf{A}(\mbf{r})\cdot(-i \nabla)$ [where the term $-i\nabla$ is the canonical electronic momentum and $\mbf{A}(\mbf{r} )$ is the vector potential of the photonic field] facilitates the photon-matter processes involving electronic resonances that usually trigger a hierarchy of multi-electronic processes, often involving nuclear rearrangements.
The remaining term $H_{\mathrm{int}}^{(2)}$ containing $\mbf{n}(\mbf{r}) \mbf{A}^2(\mbf{r} )$ [where $\mbf{n}(\mbf{r} )=\sum_{\sigma}^{}\psi_{}^\dagger(\sigma \mbf{r} ) \psi_{}^{}(\sigma \mbf{r} )$ is the electronic density operator] dominantly contributes to the scattering processes in the high-energy regime away from resonance, often taking place beyond the ionization thresholds. In the appropriate parameter regime, the role of these terms in the dynamics becomes relevant with varying degrees of importance, and consequently, the relative contributions towards the scattering cross-sections vary.
In general, the scattering events described by the Hamiltonian component $H_{\mathrm{int}}^{(2)}$ effectively take place from the electronic degrees of freedom driven by the $H_{\mathrm{int}}^{(1)}$. Often, the ionization processes triggered by the latter lead to the onset of the radiation damage \cite{lorenz2012impact}. Restriction on the fluence of the incoming sources and plausible sparsity of the distribution of the scattering centers allows for a parameter window within which the diffraction remains largely unaffected by ionization events.
We focus on the diffractive scattering processes particularly suitable for structural imaging and allow this analysis to be restricted to the scatterings induced by $H_{\mathrm{int}}^{(2)}$.
In the following, we separate the term $H_{\mathrm{int}}^{(2)}$ into initially occupied pump modes and initially unoccupied signal modes. Subsequently, as an approximation, we omit the scattering processes originating from pure pump or signal modes yielding, $\mbf{n}(\mbf{r} ) \mbf{A}^2(\mbf{r} )= (\mbf{A}_s(\mbf{r} )+\mbf{A}_p(\mbf{r} ))^2 \mbf{n}(\mbf{r} ) =(\mbf{A}_s^2(\mbf{r} )+\mbf{A}_p^2(\mbf{r} ) + 2\mbf{A}_s(\mbf{r} ) \mbf{A}_p(\mbf{r} )) \mbf{n}(\mbf{r} ) \approx 2\mbf{A}_s(\mbf{r} ) \mbf{A}_p(\mbf{r} )  \mbf{n}(\mbf{r} )$. Diffractive scattering via pure pump or signal modes may allow, as proposed in \cite{asban2019quantum}, self-heterodyning of the signal. Further, we adopt the interaction representation defined by $H_{\mathrm{int}}^{(2)}(t)= U^{\dag}(t) H_{\mathrm{int}}^{(2)} U(t)$ where $U(t) = \exp{(-i ( H_{\mathrm{matter}}+ H_{\mathrm{field}}^{}) t)} $ is the propagator.
Using the mode-expansion of the vector potential in terms of plane-waves, 
\begin{align}
    \mbf{A}(\mbf{r},t ) &= \mbf{A}_{}^{(+)}(\mbf{r},t )+\mbf{A}_{}^{(-)}(\mbf{r},t )\\
    &=\sum_{\mbf{k},\mu} \sqrt{\frac{2\pi}{V \omega_k \alpha^2}} \Big(\varepsilon_{\mbf{k},\mu} a_{\mbf{k},\mu}^{}  \exp{[i (\mbf{k}\cdot \mbf{r}-\omega_k t)]} \nonumber\\
    & + \varepsilon_{\mbf{k},\mu}^{*} a_{\mbf{k},\mu}^\dag \exp{[ -i (\mbf{k}\cdot \mbf{r}-\omega_k t)]}\Big) ,
\end{align}
and assumed rotating wave approximation, we can express the relevant Hamiltonian component as
\begin{align}\label{eq:hint2}
    H_{\mathrm{int}}^{(2)}(t)  &= 
    \alpha^2\int d\mbf{r} \sum_{\mbf{k}_s, \mbf{k}_p, \mu_s,\mu_p}  \frac{\tilde{c}_{}}{\sqrt{\omega_{\mbf{k}_s} \omega_{\mbf{k}_p}}}\lvert \varepsilon_{{\mbf{k}_s},\mu_s}^{*}\cdot \varepsilon_{{\mbf{k}_p},\mu_p}^{} \rvert \nn\\
    & a_{\mbf{k}_s,\mu_s}^{\dag} a_{\mbf{k}_p,\mu_p}^{} \Big(e^{+i \tilde{\mbf{q}}_{}\cdot \mbf{r}} \mbf{n}^{}(\mbf{r},t) \Big)e^{- i \tilde{\omega}_{\mbf{k}} t} + \mathrm{h.c.} .
\end{align}\label{eqn:hamint1}
In the above, we defined the polarization vector of the mode (indexed by $\mbf{k},\mu$) as $\varepsilon_{\mbf{k},\mu}$ and used $\tilde{c}_{} = 2\pi/V\alpha^2$, where $V$, and $ \omega_p (\omega_s)$ are the mode volume and frequency of the incoming (scattered) mode, respectively. The mode creation (annihilation) operators $a_{\mbf{k},\mu}^\dag (a_{\mbf{k},\mu}^{})$ follow the bosonic commutation relations $[a_{\mbf{k},\mu}^{},a_{\mbf{k}',\mu'}^\dag]=\delta_{\mbf{k}\mbf{k}'}\delta_{\mu \mu'}$. The difference frequency between the incoming and scattered photon modes is denoted $\tilde{\omega}_{\mbf{k}}=\omega_{\mbf{k}_{p}}-\omega_{\mbf{k}_{s}}$. In Eq.~(\ref{eq:hint2}), we note the appearance of the electronic density operator and the position-dependent phase term in the exponent, which incorporates the difference between the incoming and scattered photon momenta, $\tilde{\mbf{q}}_{}=(\mbf{k}_{p}-\mbf{k}_{s})$.\\ 
The elastic components of the diffraction signal, where the difference between the moduli of the incoming and scattered photon momenta is zero, provide structural information. In comparison, the inelastic components where the incoming and scattered photon momenta differ in magnitude contain structural and dynamical information regarding the electronic excitations. 
The applicability of elastic scattering extends to techniques such as small-angle x-ray scattering (SAXS) or wide-angle x-ray scattering (WAXS) which involves averaging over the structural disorder. They become crucial for interfaces, amorphous materials with short-range order and molecular aggregates, and proteins in the solution phase. 
\subsection{Multiphoton diffraction signal and classification}
In the high-intensity photon-matter interaction at the relevant energy regime, the typical one-pixel detection scenario monitors fewer number of modes among the full set of diffracted modes. As envisaged in Fig.~\ref{fig:diffdiag}, the scattered photonic modes would impinge on the pixelated array detector where photon intensities are registered.
The pixel-resolved photon intensity distribution, obtained by averaging over repeated measurements that monitor individual pixels separately, constitutes the diffraction pattern. We present a theoretical framework that is capable of describing the generalized diffraction and detection scenario appearing in such cases.
Towards that goal, we start by defining the diffractive scattering signal as the time-space integrated intensity of the electric field at the pixelated detector given by,
\begin{align}\label{eqn:signal00}
    &\tilde{S}_{\mrm{out}}^{(n)} = \int dt_{n} \int d\mbf{r}_{n}\cdots \int dt_{1} \int d\mbf{r}_{1} \nonumber\\&
    \ex{  E_{}^{(-)}(\mbf{r}_{n}^{},t_{n}^{}) \cdots E_{}^{(-)}(\mbf{r}_{1}^{},t_{1}^{}) E_{}^{(+)}(\mbf{r}_{1}^{},t_{1}^{}) \cdots E_{}^{(+)}(\mbf{r}_{n}^{},t_{n}^{}) }.
\end{align}
Here, the integral expression
describes the observable corresponding to the $n$-photon diffraction signal while the operator expectation is taken over the final state of the combined system consisting of the electronic matter, driving field, and detector. The signal, defined in this manner, can describe both the standard detections where the diffraction pattern is generated via averaging over several observed pixel distributions and the nonstandard situations where multi-pixel coincident detection is possible. The latter remains a forward-looking scenario due to current technological limitations. However, both cases can be described by employing different averaging procedures.
In the Keldysh-Schwinger formulation, the operator expectation value can be evaluated by accounting for the forward-backward evolution of the combined system \cite{keldysh1965diagram, fleischhauer1998quantum,  mukamel2003superoperator, dorfman2016time, dorfman2012nonlinear}.
In other words, the final state can be systematically expanded in the powers of the interaction Hamiltonian to yield different scattering configurations. Each of these configurations corresponds to a particular order of the photon-matter scattering events. Since, the operator defined above is related to the $n$-photon-matter interaction events, an expansion to the $n$-th order in $H_{\mathrm{int}}^{(2)}(t)$ for the bra and ket generates the desired signal. A compact expression for the signal can be obtained by collecting individual path-ordered operators and averaging them over the initially correlated combined state $\phi_{\mathrm{in}}^{}$. Furthermore, if a factorizibility assumption on the initially correlated state is made, it implies $ \phi_{\mathrm{in}}^{} \equiv \phi_{\mrm{matter},\mrm{in}}\otimes \phi_{\mrm{field},\mrm{in}} $ where the term $\phi_{\mrm{matter},\mrm{in}}$ represents the state of the correlated electronic matter prior to the scattering events, and $\phi_{\mrm{field},\mrm{in}}$ represents the initial state of the incoming photonic sources. For the latter, it has been assumed that the scattered modes are in a vacuum while the rest of the modes are in an arbitrary field state. The prior choice of neglecting $H_{\mathrm{int}}^{(1)}$ during the expansion signifies the absence of externally induced electronic current. The lack of nontrivial field-induced correlation among the electronic modes 
justifies the factorized form of the initial state.
\begin{figure*}[ht]
\includegraphics[width=\textwidth]{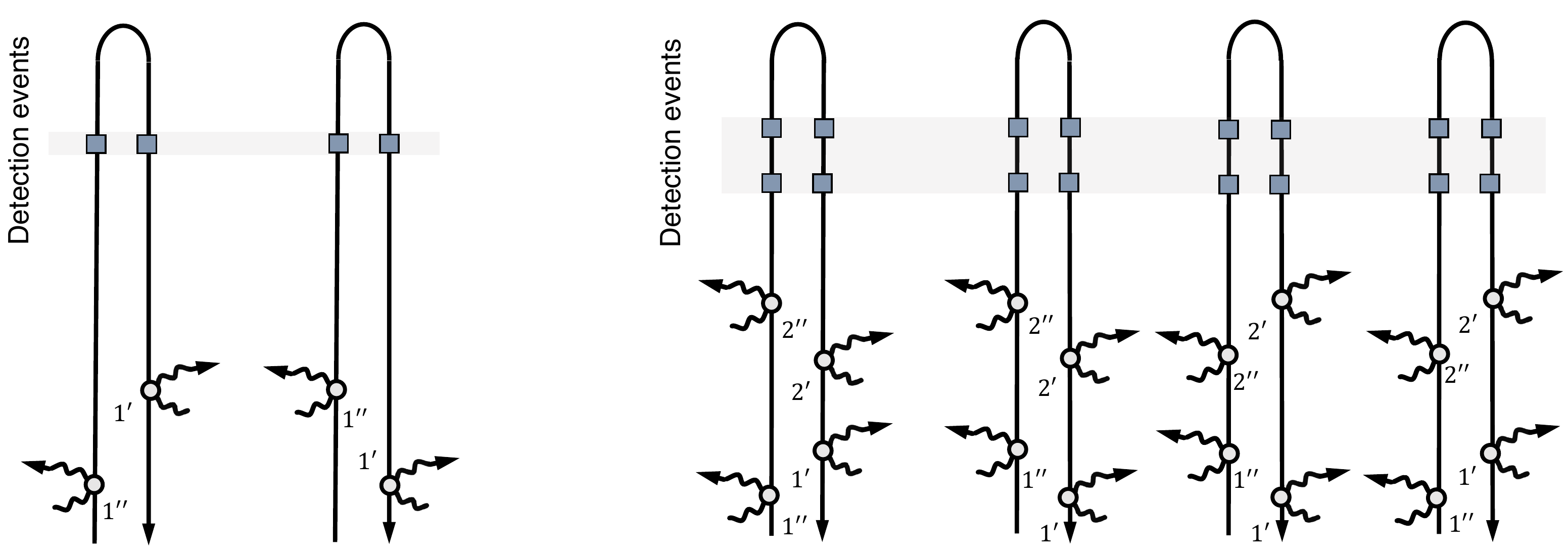}
\caption{Diagrams describing one-photon (left) and two-photon (right) diffractive scattering events. The highlighted patch, in gray, denotes the detection process. The space-time arguments are abbreviated as $1' \equiv (\mbf{r}_1',t_1')$. Individual diagrams signify distinct pathways which make interfering contributions to the observed signal. The number of pathways is two and four, respectively, for the above-mentioned scattering events. The pathway contributions may be manipulated via shaping of the incoming photonic modes and detection configurations.} 
\label{fig:supdiag12}
\end{figure*}
In this limit, the operator expectation value defined in Eq.~(\ref{eqn:signal00}) can be expressed as
\begin{widetext}
\begin{align}
\label{eqn:sigbarefull0}
 & \tilde{S}_{\mrm{out}}^{(n)} = 
 \int dt_{n} \int d\mbf{r}_{n} \cdots \int dt_{1} \int d\mbf{r}_{1} \times
\int d\mbf{r}_{n''}\int dt_{n''}  \cdots
\int_{}^{}  d\mbf{r}_{1''}\int dt_{1''} \times
\int d\mbf{r}_{n'} dt_{n'} \cdots \int_{}^{} d\mbf{r}_{1'} dt_{1'} \nonumber\\&
\qquad D_{s}^{(n)}(\mbf{r}_n, t_n, \cdots \mbf{r}_1, t_1 ; \mbf{r}_{1'}, t_{1'} \cdots \mbf{r}_{n'}, t_{n'};  \mbf{r}_{n''}, t_{n''} \cdots  \mbf{r}_{1''}, t_{1''}) \nonumber\\&
\qquad K_{}^{(n)}( \mbf{r}_{1'}, t_{1'} \cdots \mbf{r}_{n'}, t_{n'};  \mbf{r}_{n''}, t_{n''} \cdots \mbf{r}_{1''}, t_{1''})
 \tilde{D}_p^{(2)}( \mbf{r}_{1'}, t_{1'} \cdots \mbf{r}_{n'}, t_{n'};  \mbf{r}_{n''}, t_{n''} \cdots  \mbf{r}_{1''}, t_{1''}) .
\end{align}
\end{widetext}
In the above, the arguments of the operators have been distinguished to indicate the time and path ordering. In the expression, we introduced a set of multipoint, space-time dependent correlation functions for the sake of brevity. In particular, we defined the correlation function of the scattered photon and detector modes. It can be expressed as the sum of terms consisting of the products of elementary two-point correlation functions given by
\begin{align}\label{eqn:sigcorr}
& D_{s}^{(n)}(\mbf{r}_n, t_n, \cdots \mbf{r}_1, t_1 ; \mbf{r}_{1'}, t_{1'} \cdots \mbf{r}_{n'}, t_{n'};  \mbf{r}_{n''}, t_{n''} \cdots  \mbf{r}_{1''}, t_{1''}) \nonumber\\&
    =\sum_{\{m\}} \prod_{j',j'',d}
  D_{s,m}^{(1)}(\mbf{r}_{d}^{},t_d^{}; \mbf{r}_{j'}^{},t_{j'}^{})
 D_{s,m}^{(1),*}(\mbf{r}_{d}^{},t_d; \mbf{r}_{j''}^{}, t_{j''}^{}).
\end{align}
Here the subscript $d \in \{1 \cdots n\}$ and $j',j''\in \{1' \cdots n' ; 1'' \cdots n''\}$ are associated with the detection and scattering events, respectively. The elementary two-point correlation functions are expressed as 
\begin{align}
     D_{s}^{(1)}(\mbf{r}_{d}^{}, t_{d}^{};\mbf{r}_{j}^{}, t_{j}^{}) 
    = \ex{
\mbf{E}_{d}^{(+)}(\mbf{r}_{d}^{}, t_{d}^{}) \mbf{A}_{s}^{(-)}(\mbf{r}_{j}^{}, t_{j}^{})} .
\end{align}
These functions are generated by considering $2$-tuples formed by one scattering and one detection index lying within a particular branch. Two such pairings from the separate branches yield a correlation function pair, each of which corresponds to the population of one quantum in the detection field mode. The summation over the combinatorial set $\{m\}$ in Eq.~(\ref{eqn:sigcorr}) contains products over such correlation function pairs.
The contraction of a total of $2n$ scattering indices with the $2n$ detection indices while keeping the branch specificity yields distinct detection configurations. The detection configurations represent pathways that interfere among themselves to give rise to the observed diffraction pattern upon averaging.
\begin{figure*}[ht]
\includegraphics[width=\textwidth]{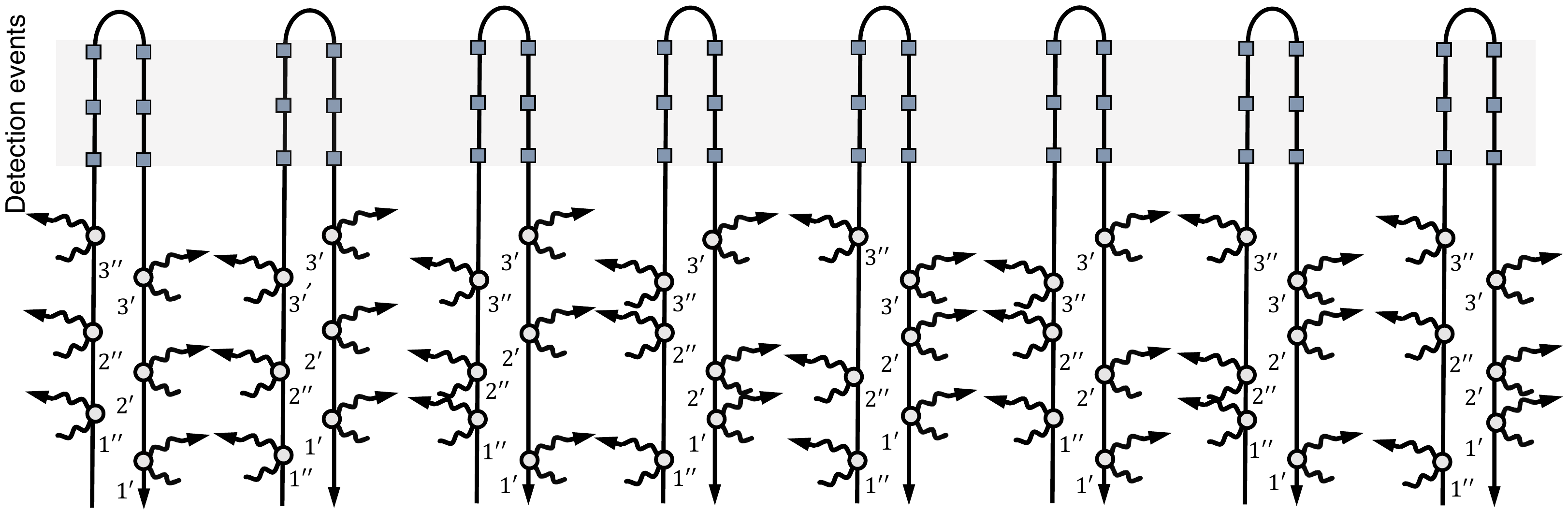}
\caption{Diagram describing three-photon diffractive scattering event. The highlighted patch and the abbreviations are defined as in Fig.~\ref{fig:supdiag12}. The number of pathways in this case is eight.
} 
\label{fig:supdiag03}
\end{figure*}
For the cases of one- and two-photon diffractive scattering, diagrams in Figs.~\ref{fig:supdiag12} and \ref{fig:supdiag03} provide corroborative illustrations, while Eqs.~(\ref{eqn:AppDs2}) and (\ref{eqn:AppDs3}) provide the relevant expressions. We note that even when all the detection times are set as equal, the statistical weights coming from various detection configurations should be taken into account while evaluating the signal.
In Eq.~(\ref{eqn:sigbarefull0}), we also defined the incoming photonic field correlation function, which can be expressed as
\begin{align}\label{eqn:fieldcorr}
   & \tilde{D}_p^{(n)}( \mbf{r}_{1'}, t_{1'} \cdots \mbf{r}_{n'}, t_{n'};  \mbf{r}_{n''}, t_{n''} \cdots \mbf{r}_{1''}, t_{1''})\nonumber\\&
   =\ex{  
   \mbf{A}_{p}^{(-)}(\mbf{r}_{1'},t_{1'}) \cdots 
   \mbf{A}_{p}^{(-)}(\mbf{r}_{n'},t_{n'}) \nonumber\\&
  \qquad \times\mbf{A}_{p}^{(+)}(\mbf{r}_{n''}, t_{n''})\cdots \mbf{A}_{p}^{(+)}(\mbf{r}_{1''},t_{1''}) }.
\end{align}
The spatio-temporal coherence and statistical properties of the incoming photonic field encoded in this term dictate the nature of the correlation between the scattering events \cite{vartanyants2010coherence, pietsch2005coherence, skopintsev2014characterization, khubbutdinov2019coherence}.
Hence, a prior characterization of its functional nature is essential for carrying out controlled diffractive scattering as well as a reliable posterior data inversion. Furthermore, we defined the electronic density correlation function 
composed of the space-time dependent density operators $\mbf{n}(\mbf{r}_{}, t_{})$ expressed as
\begin{align}
    \label{eqn:matcorr}
& K_{}^{(n)}( \mbf{r}_{1'}, t_{1'} \cdots \mbf{r}_{n'}, t_{n'};  \mbf{r}_{n''}, t_{n''} \cdots \mbf{r}_{1''}, t_{1''})
\nonumber\\
&= \ex{  
   \mbf{n}_{}^{}(\mbf{r}_{1'},t_{1'})\cdots
   \mbf{n}_{}^{}(\mbf{r}_{n'},t_{n'}) \nonumber\\&
  \qquad \times\mbf{n}_{}^{}(\mbf{r}_{n''}, t_{n''})\cdots \mbf{n}_{}^{}(\mbf{r}_{1''},t_{1''}) }.
\end{align}
Since at this stage, the electronic density operators do not assume any ad hoc separation in terms of a noninteracting and a correlated part, in principle, this function retains the full set of information regarding the correlation and fluctuation properties of the electronic degrees of freedom \cite{van1954correlations}.\\
The expression in Eqs.~(\ref{eqn:sigbarefull0}) captures, in essence, the phenomenology of multiphoton diffraction by making the role of the density correlations, the incoming photonic field properties, and the outgoing scattered field configurations explicit. The electronic density correlations carry information regarding the relative spatial distribution of scattering centers. In principle, the scattered photonic modes have signatures originating from both independent and correlated scattering events. The relevant information about these events is encoded in the specific path and time ordering of the space-time arguments of the correlation functions. The fact that several such scattering events jointly contribute to the signal mode being populated is made evident via the convolutional nature of the expression.
The role of the incoming photonic field can be thought of as the one facilitating the manipulation of the electronic spectral weights of the bare system. Consequently, the incoming photonic sources and the detection schemes can be jointly manipulated for carrying out controlled diffraction measurements aimed at suppressing or amplifying specific signal components.
\section{Resources and factorizability constraints for correlated diffraction}
The expression in Eq.~(\ref{eqn:sigbarefull0}) identified the role of three correlation resources arising from the configuration of the scattered modes [Eq.~(\ref{eqn:sigcorr})], the correlation properties of the incoming photonic modes [Eq.~(\ref{eqn:fieldcorr})], and the intrinsic correlations of the electronic matter [Eq.~(\ref{eqn:matcorr})].
The nontrivial mode correlations manifest themselves in terms of the nonfactorizability of correlation functions in terms of the lower order ones. In this section, we systematically explore the factorization properties of correlation functions corresponding to the above-mentioned resources. In particular, we discuss the nature of the dynamical inter-mode correlation in each case and analyze, in physical terms, the assumptions required for their factorization. For this purpose, we begin by considering the convolutional expression of the signal detection given in Eq.~(\ref{eqn:sigbarefull0}).
\subsection{Role of factorizable scattered field correlation function}
The observability of the signatures of correlations in the diffractive scattering pattern depends crucially on the details of the detection scheme. In mathematical terms, the latter involves a convolution of the scattered photonic modes with the detector response function. Typically, the detector response is a functional of the probability of detecting photonic modes within a specified energy window and the pixel distribution function.  
The only source of statistical correlation in Eq.~(\ref{eqn:sigbarefull0}) has been incorporated by considering the sum of combinatorial terms containing the products of correlation function pairs given in Eq.~(\ref{eqn:sigcorr}). The expression entails the fact that the final state interactions among the scattered modes are neglected. The number of terms under the summation denotes the number of ways in which the outgoing scattered field modes can be combined with the predesignated detection field modes. The corresponding expression for the two- and three-photon diffractive scattering events, given in Eqs.~(\ref{eqn:AppDs2}) and (\ref{eqn:AppDs3}), shows various correlation functions related to distinct detection pathways.
Given a physical scenario where the transverse coherence of the incoming photonic source is sufficiently high and the multipoint electronic correlation function does not trivially factorize, the spectral weights associated with these statistical contributions solely determine the resolution of the correlation-induced features.
\begin{figure*}[ht]
\includegraphics[width=\textwidth]{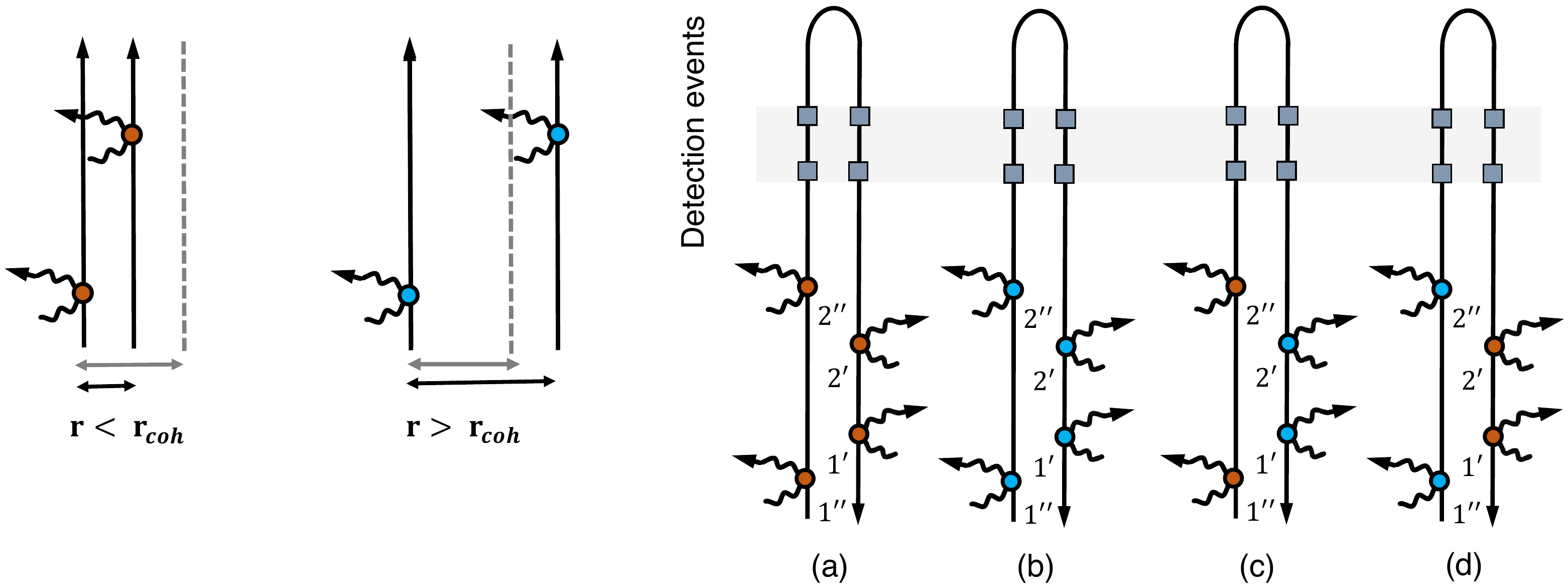}
\caption{Diagrams describing the classification of one particular diagram from the suite of two-photon diffractive scattering events for spatially distributed scattering (indicated on the left). The scattering events that are spatially close (i.e., at length scales shorter than the coherence length) have been denoted by the orange-colored vertex, and the ones that are spatially far apart (i.e., at length scales larger than the coherence length) are denoted by the blue-colored vertex. 
These lead to different subclasses of diagrams.} 
\label{fig:supdiag2space}
\end{figure*}
\subsection{Role of factorizable incoming photonic field correlation function}
In certain cases, the spatio-temporal coherence properties of the incoming photonic sources become the dominant governing factor that determines the nature of correlated scattering processes. Typically, the incoming photonic modes are scattered from the distribution of scattering centers. Information regarding the time-dependent relative spatial distribution of the scattering centers is contained in the correlation and fluctuation characteristics of the electronic densities. The latter alters the statistics of the scattered photonic modes and gives rise to diagnostic interference patterns at the detector. 
If the temporal and transverse coherence lengths of the incoming photonic modes are shorter than the spatio-temporal coherence lengths associated with the internal electronic processes, the multipoint incoming field correlation function can be represented in a factorized form without loss of generality. In its simplest form, the factorized form can be expressed in terms of a sum containing products of two-point correlation functions as given below
\begin{align}\label{eqn:fieldcorrfac}
   & \tilde{D}_{p}^{(n)}( \mbf{r}_{1'}, t_{1'}\cdots \mbf{r}_{n'}, t_{n'};  \mbf{r}_{n''}\cdots t_{n''},  \mbf{r}_{1''}, t_{1''})\nonumber\\&
   =\sum_{\{w\}} \prod_{j',j''}^{} \tilde{D}_{p,w}^{(1)}(\mbf{r}_{j'}^{},t_{j'}^{} ; \mbf{r}_{j''}^{},t_{j''}^{} ) .
\end{align}
Here, the terms under the summation from a combinatorial set
constructed by taking $2$-tuples constituted of indices $\{j',j''\}$. The product of such pairings forms a unique combinatorial element, denoted by $w$. For two- and three-photon diffractive scattering, these elements can be easily obtained by pairing vertices lying in separate branches of the corresponding diagrams. The expressions in Eqs.~(\ref{eqn:AppDp2}) and (\ref{eqn:AppDp3}) give their analytical forms, respectively.
Using Eq.~(\ref{eqn:fieldcorrfac}), the operator expectation value in Eq.~(\ref{eqn:sigbarefull0}) assumes a simpler form
\begin{widetext}
\begin{align}\label{eqn:sigbarefull01}
  \tilde{S}_{\mrm{out},1}^{(n)} &=
 \int dt_{n} d\mbf{r}_{n} \cdots \int dt_{1} d\mbf{r}_{1} 
\int d\mbf{r}_{n''} dt_{n''}  \cdots
\int_{}^{}  d\mbf{r}_{1''} dt_{1''}
\int d\mbf{r}_{n'} dt_{n'} \cdots \int_{}^{} d\mbf{r}_{1'} dt_{1'} \nonumber\\&
\times D_{s}^{(n)}(\mbf{r}_n, t_n, \cdots \mbf{r}_1, t_1 ; \mbf{r}_{1'}, t_{1'} \cdots \mbf{r}_{n'}, t_{n'};  \mbf{r}_{n''}, t_{n''} \cdots  \mbf{r}_{1''}, t_{1''})  \nonumber\\&
\qquad K_{}^{(n)}( \mbf{r}_{1'}, t_{1'} \cdots \mbf{r}_{n'}, t_{n'};  \mbf{r}_{n''}, t_{n''} \cdots \mbf{r}_{1''}, t_{1''})
\Big\{\sum_{\{w\}} \prod_{j',j''}^{} \tilde{D}_{p,w}^{(1)}(\mbf{r}_{j'}^{},t_{j'}^{} ; \mbf{r}_{j''}^{},t_{j''}^{} )\Big\}.
\end{align} 
\end{widetext}
In the above, due to the spatial integrations involved, the domain of values of the incoming field correlation functions imposes physical bounds on the nature of the electronic density correlations.
Typically, the scattering events at higher energies involve large momentum transfers between the incoming and scattered photons. It restricts the long-range electronic correlations from developing and leads to shorter coherence length scales within the observational window.
If the coherence length scales associated with the density correlations are within the bound, it may lead to separation and conditional factorization of the multidimensional integral.
The physical argument based on the properties of the density correlation functions that may allow a semi-quantitative analysis will be presented in the next section.
In a separate scenario, where the out-of-equilibrium density fluctuations are present, the scattering events at a certain instance may be taken as a source for the subsequent scattering events. In such cases, scattering events would experience dynamically cumulative effects exerted by the preceding ones. The present expression is capable of describing such situations arising from static and dynamic correlated density fluctuations via appropriate modifications of the operator expectation value.
In multiphoton interactions as well as in cases where the photonic source consists of several sub-pulses, such effects may arise.
We note that the knowledge regarding plausible factorizability of the incoming field correlation function is also essential for carrying out meaningful reconstruction.
\subsection{Role of factorizable electronic density correlation function} 
As made explicit in Eq.~(\ref{eqn:sigbarefull0}), for a chosen incoming field correlation function and detection mode configuration, the features related to correlation in the diffractive scattering signal can be interpreted by invoking the many-body nature of the electronic modes. In fact, the resources mentioned in the previous subsections are made relevant (irrelevant) by the presence (absence) of electronic correlations in the underlying matter.
The electronic correlations relevant for structural imaging may have various physical origins, ranging from Coulomb-mediated electronic interactions to phonon-induced renormalization present in the ground state. In this analysis, the explicit contributions due to the induced electronic current have not been treated and corresponding scenarios involving electronic resonances e.g., higher-order dispersive effects, and electron-electron scattering are omitted.
To explore the conditions that may lead to the plausible factorization of the electronic correlation functions, we focus on the previously introduced $n$-point density correlation function, in Eq.~(\ref{eqn:matcorr}). This expression can not be reexpressed in terms of the lower-order factorized counterparts due to the many-particle nature of the function. However, exceptions may arise due to the physically motivated assumptions leading to the separation of space-time variables. One such condition may originate while considering the joint spatio-temporal characteristics of the incoming field and electronic correlation properties.
To describe such a scenario, we introduce the notion of material-specific spatio-temporal coherence length and define the corresponding parametric variable 
$\lambda(\mbf{r}_{1} \cdots \mbf{r}_{n},t_{1} \cdots t_{n})$, which is a function of the distribution of
scattering centers in the real space.
Using the definition, without any loss of generality, we may recast the expression for the density correlation function as
\begin{align}\label{eqn:matcorrfac}
& K_{}^{(n)}( \mbf{r}_{1'}, t_{1'} \cdots \mbf{r}_{n'}, t_{n'};  \mbf{r}_{n''}, t_{n''} \cdots  \mbf{r}_{1''}, t_{1''}) \approx \sum_{ \{g\}} \prod_{j',j''}\nonumber\\&
   K_{0,g}^{(1)}(\mbf{r}_{j'}^{},t_{j'}^{},\mbf{r}_{j''}^{}, t_{j''}^{})      \exp\{{-\lambda_g(\mbf{r}_{j'}^{},t_{j'}^{},\mbf{r}_{j''}^{},t_{j''}^{})\}} ,
\end{align}
where $ K_{0,g}^{(1)}(\mbf{r}_{j'}^{},t_{j'}^{},\mbf{r}_{j''}^{}, t_{j''}^{})$ is the reference density correlation function defined in the limit where the multipoint density correlation function factorizes. The associated exponent defines a cumulant functional dependent on the coherence length-scale parameters which, in principle, incorporates the space-time dependent effects arising from electronic correlation. The set of combinatorial terms is constituted by considering all possible pairings involving the $\{j',j''\}$ indices. A particular choice of all such pairings from the set of $2n$ indices is denoted $g$. Notably, the nature of the cumulant functional is also dependent on particular elements of the combinatorial set.
Depending on the physical scenario, the cumulant functional can be used to define spatial and temporal scaling properties of the electronic density correlations. Corresponding scaling parameters would depend on conditions such as the nature of itinerant electron-electron interaction-induced correlations governing localization properties of the electronic wave functions, the interatomic potentials, and the nature of the disorders. The latter two features are particularly important for liquids and amorphous materials, respectively. In the absence of significant inter-particle interactions, the coherence lengths become comparatively short-ranged.
As a consequence, the cumulant function may decay faster than the reference density correlation function. In such cases, we would replace the cumulant function with a $c$-number and arrive at a partially factorized limit. In general, any physical assumption that lead to the systematic separation of multipoint space-time variables would result in the factorization of the electronic correlation functions in terms of their lower-order counterparts. A diagrammatic interpretation that may guide such intuitive length scale separation has been described in Fig.~\ref{fig:supdiag2space}. Therein, a classification scheme based on whether the separation between the scattering centers is larger (shorter) than the coherence length scale parameter is shown to generate several subclasses of diagrams for one particular configuration of the detection and incoming field. Furthermore, in cases where the properties of the incoming field correlation allows separation of the integral variables into non-interacting space-time clusters, the multipoint electronic correlation functions undergo factorization that is dominantly governed by the same.
It leads to the field and electronic density correlation functions being associated with the same set of combinatorial indices, $w$. Furthermore, due to the weak presence of correlation in the incoming field, the cumulant functional in the signal expression can be trivially set to unity, which yields the completely factorizable limit. In that limit, the reference density correlation function can be presented as a sum of the products of renormalized two-point density correlation functions. The terms that form the relevant combinatorial set are explicitly given in Eqs.~(\ref{eqn:AppK2}) and (\ref{eqn:AppK3}). These terms and the ones given in Eqs.~(\ref{eqn:AppKaux2}) and (\ref{eqn:AppKaux3}) put together would constitute the larger subset $\{g\}$ indicated in Eq.~(\ref{eqn:matcorrfac}).
Incorporating the assumptions stated above, we recover the signal given in Eq.~(\ref{eqn:sigbarefull0}) in the factorized limit as
\begin{widetext}
\begin{align}\label{eqn:approx2facline1}
\tilde{S}_{\mrm{out},2}^{(n)} &= 
\int dt_{n} d\mbf{r}_{n} \cdots \int dt_{1} d\mbf{r}_{1} 
\int d\mbf{r}_{n''} dt_{n''}  \cdots
\int_{}^{}  d\mbf{r}_{1''} dt_{1''}
\int d\mbf{r}_{n'} dt_{n'} \cdots \int_{}^{} d\mbf{r}_{1'} dt_{1'} \nonumber\\&
\times D_{s}^{(n)}(\mbf{r}_n, t_n, \cdots \mbf{r}_1, t_1 ; \mbf{r}_{1'}, t_{1'} \cdots \mbf{r}_{n'}, t_{n'};  \mbf{r}_{n''}, t_{n''} \cdots  \mbf{r}_{1''}, t_{1''}) 
\nonumber\\&
\qquad \Big\{\sum_{\{w\}} \prod_{j',j''}^{}
K_{0,w}^{(1)}(\mbf{r}_{j'}^{},t_{j'}^{},\mbf{r}_{j''}^{}, t_{j''}^{}) \sum_{\{w\}} \prod_{j',j''}^{}
\tilde{D}_{p,w}^{(1)}(\mbf{r}_{j'}^{},t_{j'}^{} ; \mbf{r}_{j''}^{},t_{j''}^{} )\Big\}  .
\end{align} 
\end{widetext}
Even though the signal expression is evaluated in the factorized limit, we note that the correlation contributions are combinatorial in nature. The information remains encoded into the various products of the $2$-point correlation functions involving electronic densities, incoming and scattered field operators.
In order to make the role of the involved photonic field modes explicit, we insert the mode-decomposition of the incoming, scattered, and detected field modes in Eq.~(\ref{eqn:approx2facline1}) and recast the signal expression as
\begin{widetext}
\begin{align}\label{eqn:sigbarefull}
&  S_{\mrm{out},2}^{(n)} =(
\alpha^{4})^n
\int dt_{n} d\mbf{r}_{n} \cdots \int dt_{1} d\mbf{r}_{1} 
\int d\mbf{r}_{n''} dt_{n''}  \cdots
\int_{}^{}  d\mbf{r}_{1''} dt_{1''}
\int d\mbf{r}_{n'} dt_{n'} \cdots \int_{}^{} d\mbf{r}_{1'} dt_{1'}  
\sum_{\substack{\{\mu_{s_{j}}, \mu_{p_{j}}\}\\
\{\mbf{k}_{s_{j}}, \mbf{k}_{p_{j}}\} }}
(\tilde{c}_{}^{2} )^{n}\nonumber\\&
\frac{ \varepsilon_{{\mbf{k}_{s_{1'}}},\mu_{s_{1'}}}^{}\cdot  \varepsilon_{{\mbf{k}_{p_{1'}}},\mu_{p_{1'}}}^{*}   }{\sqrt{\omega_{\mbf{k}_{s_{1'}}} \omega_{\mbf{k}_{p_{1'}}}}} 
\cdots
\frac{ \varepsilon_{{\mbf{k}_{s_{n'}}},\mu_{s_{n'}}}^{}\cdot  \varepsilon_{{\mbf{k}_{p_{n'}}},\mu_{p_{n'}}}^{*}  }{\sqrt{\omega_{\mbf{k}_{s_{n'}}} \omega_{\mbf{k}_{p_{n'}}}}}
\frac{ \varepsilon_{{\mbf{k}_{s_{n''}}},\mu_{s_{n''}}}^{*}\cdot  \varepsilon_{{\mbf{k}_{p_{n''}}},\mu_{p_{n''}}}^{} }{\sqrt{\omega_{\mbf{k}_{s_{n''}}} \omega_{\mbf{k}_{p_{n''}}}}} 
\cdots
\frac{ \varepsilon_{{\mbf{k}_{s_{1''}}},\mu_{s_{1''}}}^{*}\cdot  \varepsilon_{{\mbf{k}_{p_{1''}}},\mu_{p_{1''}}}^{} }{\sqrt{\omega_{\mbf{k}_{s_{1''}}} \omega_{\mbf{k}_{p_{1''}}}}} \nonumber\\&
\Big\{
\sum_{\{m\} } \prod_{j',j''}\prod_{d}
   \ex{a_{\mbf{k}_{s_{j'}},\mu_{s_{j'}}}^{}
a_{\mbf{k}_{d},\mu_{d}}^{\dag}} \ex{a_{\mbf{k}_{d},\mu_{d}}^{}
a_{\mbf{k}_{s_{j''}},\mu_{s_{j''}}}^{\dag}} \Big\}
\nonumber\\&
\Big\{\sum_{\{w\} }  \prod_{j',j''=1}^{n} K_{0,w}^{(1)}(\{\mbf{r}_{j'}^{},t_{j'}^{},\mbf{r}_{j''}^{},t_{j''}^{}\})  \Big\}
\exp({-i \tilde{\mbf{q}}_{1'}^{} \cdot\mbf{r}_{1'}^{}
\cdots 
-i \tilde{\mbf{q}}_{n'}^{} \cdot\mbf{r}_{n'}^{}
+i \tilde{\mbf{q}}_{n''}^{} \cdot\mbf{r}_{n''}^{}
\cdots 
+i \tilde{\mbf{q}}_{1''}^{} \cdot\mbf{r}_{1''}^{} )}
\nonumber\\&
\Big\{\sum_{\{w\} }  \prod_{j',j''=1}^{n} 
\ex{
a_{\mbf{k}_{p_{j'}},\mu_{p_{j'}}}^{\dag} a_{\mbf{k}_{p_{j''}},\mu_{p_{j''}}}^{}} \Big\}
\exp({
+i \tilde{\omega}_{\mbf{k}_{1'}}^{} t_{1'}^{}
\cdots 
+i \tilde{\omega}_{\mbf{k}_{n'}}^{} t_{n'}^{}
-i \tilde{\omega}_{\mbf{k}_{n''}}^{} t_{n''}^{}
\cdots
-i \tilde{\omega}_{\mbf{k}_{1''}}^{} t_{1''}^{}} ),
\end{align}
\end{widetext}
where we used the notation $\{\mu_{s_{j}}, \mu_{p_{j}}\} $  and $\{\mbf{k}_{s_{j}}, \mbf{k}_{p_{j}}\} $ to indicate the set of polarization values $ \mu_{s_{1}}\cdots \mu_{s_{n}}, \mu_{p_{1}} \cdots \mu_{p_{n}}$ and momentum values $ \mbf{k}_{s_{1}}\cdots \mbf{k}_{s_{n}}, \mbf{k}_{p_{1}} \cdots \mbf{k}_{p_{n}} $ lying in either of the branches over which the summation would be carried out.
We note that the factorized correlation functions have common exponential factors along with permutable prefactors given inside the bracketed expressions.
The electronic density correlation functions are weighted by the exponential factors originating from the incoming and outgoing modes (i.e., $e^{-i \tilde{\mbf{q}}_{j}\cdot \mbf{r}_{j}}$), which incorporate effective transferred photon momentum. During the spatial integration in Eq.~(\ref{eqn:sigbarefull}), these position and momentum-dependent phase terms determine the range in the real space over which the density correlation functions acquire significant values. In other words, in the limit of large transferred momentum, these weighting terms govern the nature of the spatial integration and help identify the localized features in the real space.
The expression, although derived in the factorized limit, still presents a generalized scenario capable of describing both elastic and inelastic scattering. The correlator involving detection and scattered mode operators which describes the detection scheme are usually evaluated by vacuum averaging.
\subsection{Limiting expressions and statistical estimators}
This subsection is intended to describe another set of assumptions that are necessary for recovering the simplest generalization of commonly used diffractive scattering signal. 
For this purpose, we develop the expression given in Eq.~(\ref{eqn:approx2facline1}) further for a source consisting of a coherent ensemble of pulses, and assuming a simpler detection configuration.
Motivated by the argument that the signal is dominated by the effects arising from the incoming field intensities and, the intensity correlation between the distant space-time indices does not exert significant influence, we approximate the electronic density correlation function in Eq.~(\ref{eqn:matcorrfac}) by keeping the symmetrically paired indices.
It leads to the form,
\begin{align}\label{eqn:matcorrfac1}
&  K_{}^{(n)}( \mbf{r}_{1'}, t_{1'}\cdots \mbf{r}_{n'}, t_{n'};  \mbf{r}_{n''}, t_{n''} \cdots  \mbf{r}_{1''}, t_{1''}) 
\approx
 \nonumber\\&
K_{0}^{(1)}(\mbf{r}_{1'}^{},t_{1'}^{} ;\mbf{r}_{1''}^{},t_{1''}^{}) \cdots K_{0}^{(1)}(\mbf{r}_{n'}^{},t_{n'}^{} ;\mbf{r}_{n''}^{},t_{n''}^{}) .
\end{align}
Moreover, we consider a situation where the bandwidth of the incoming photonic source encompasses the energy range of the electronic dynamics of interest.
Further, assuming that the envelope of the incoming pulses has a narrow bandwidth and a smaller angular spread, it leads to a case where the largest spectral weights of the field modes are around a mean wavevector $\mbf{k}_{p_{j}}$ and, polarization $\mu_{p_{j}}$. It allows the replacements,
$\sqrt{\omega_{\mbf{k}_{p_{j'}}} \omega_{\mbf{k}_{p_{j''}}}} \approx \omega_{\mbf{k}_{p_{j}}}$ and $\varepsilon_{{\mbf{k}_{p_{j'}}},\mu_{p_{j'}}}^{} = \varepsilon_{{\mbf{k}_{p_{j''}}},\mu_{p_{j''}}}^{} \approx \varepsilon_{{\mbf{k}_{p_{j}}},\mu_{p_{j}}}^{}$. 
If the incoming pulses have negligible spatial dependence compared to the material specific coherence length scales, a replacement of the position dependence of the field profile by its value at $\mbf{r}_0$ can be made. This leads to the factorization of the field intensities, keeping only the terms which are commensurate with the approximation on electronic density operators. We also assume, $\omega_{\mbf{k}_{s_{j'}}} =\omega_{\mbf{k}_{s_{j''}}} \approx \omega_{\mbf{k}_{s_{j}}}$ and $\varepsilon_{{\mbf{k}_{s_{j'}}},\mu_{s_{j'}}}^{} = \varepsilon_{{\mbf{k}_{s_{j''}}},\mu_{s_{j''}}}^{} \approx \varepsilon_{{\mbf{k}_{s_{j}}},\mu_{s_{j}}}^{}$. Alongside, it is convenient to move to a new set of time variables defined as $\bar{t}_j=(t_{j''}+t_{j'})/2$ and $\bar{\tau}_j=(t_{j''}-t_{j'})$. 
Finally, we arrive at the expression
\begin{widetext}
\begin{align}\label{eqn:sigbarefullfac2}
   &  S_{\mathrm{out},3}^{(n)} 
 = (\alpha^{4})^n
\int dt_{n} d\mbf{r}_{n} \cdots \int dt_{1} d\mbf{r}_{1}
 \int d\mbf{r}_{n''} \cdots d\mbf{r}_{1''} 
\int d\mbf{r}_{n'} \cdots d\mbf{r}_{1'} \times
\int_{}^{} d\bar{t}_{n} d\bar{\tau}_{n} \cdots d\bar{t}_{1} d\bar{\tau}_{1}  \nonumber\\& 
\sum_{\substack{\{\mu_{s_{j}}, \mu_{p_{j}}\}\\
\{\mbf{k}_{s_{j}}, \mbf{k}_{p_{j}}\} }}
\tilde{c}_{}^{2n} 
\Bigg(\frac{\lvert \varepsilon_{{\mbf{k}_{s_{1}}},\mu_{s_{1}}}^{*}\cdot  \varepsilon_{{\mbf{k}_{p_{1}}},\mu_{p_{1}}}^{} \rvert  }{\sqrt{\omega_{\mbf{k}_{s_{1}}} \omega_{\mbf{k}_{p_{1}}}}} \cdots
\frac{\lvert \varepsilon_{{\mbf{k}_{s_{n}}},\mu_{s_{n}}}^{*}\cdot  \varepsilon_{{\mbf{k}_{p_{n}}},\mu_{p_{n}}}^{} \rvert  }{\sqrt{\omega_{\mbf{k}_{s_{n}}} \omega_{\mbf{k}_{p_{n}}}}} \Bigg)^2
\Big\{\sum_{\{m\} } \prod_{d,j',j''}
\delta_{\mbf{k}_{s_{j'}}\mbf{k}_{d}}\delta_{\mu_{s_{j'}}\mu_{d}} 
\delta_{\mbf{k}_{s_{j''}}\mbf{k}_{d}}\delta_{\mu_{s_{j''}}\mu_{d}} \Big\}
\nonumber\\&
\qquad \Big\{ K_{0}^{(1)}(\mbf{r}_{1'}^{},t_{1'}^{} ;\mbf{r}_{1''}^{},t_{1''}^{}) \cdots K_{0}^{(1)}(\mbf{r}_{n'}^{},t_{n'}^{} ;\mbf{r}_{n''}^{},t_{n''}^{}) \Big\}
\exp({-i \tilde{\mbf{q}}_{1'}^{} \cdot\mbf{r}_{1'}^{}
\cdots 
-i \tilde{\mbf{q}}_{n'}^{} \cdot\mbf{r}_{n'}^{}
+i \tilde{\mbf{q}}_{n''}^{} \cdot\mbf{r}_{n''}^{}
\cdots 
+i \tilde{\mbf{q}}_{1''}^{} \cdot\mbf{r}_{1''}^{} )}
\nonumber\\&
\qquad \Big\{ I_{}^{}(\mbf{r}_{0},\bar{\tau}_{1},\bar{t}_{1} ) \cdots I_{}^{}(\mbf{r}_{0},\bar{\tau}_{n},\bar{t}_{n} ) \Big\}
\exp({
-i \tilde{\omega}_{\mbf{k}_{1}}^{} \bar{\tau}_{1}^{}
\cdots 
-i \tilde{\omega}_{\mbf{k}_{n}}^{} \bar{\tau}_{n}^{}
} ).
\end{align}
\end{widetext}
In the above, $I(\mbf{r}_0,\bar{\tau}_{j},\bar{t}_{j})$ terms denote the incoming field intensities. In general, purely classical, nonstatistical states of the incoming field would give rise to a situation where the combinatorial set containing two-point electronic density correlation functions are multiplied by the classical amplitude functions of the incoming field.
Eq.~(\ref{eqn:sigbarefullfac2}) is a generalization of conclusions obtained in previous works which, for one-photon diffractive scattering, stated that the signal depends on the electronic correlation functions.
For the case of Gaussian pulse profiles, exploiting the separation of temporal variables, the time integrals can be recast into two independent ones and the expression can be rewritten in the following form
\begin{widetext}
\begin{align}\label{eqn:sigbarefullfac3}
   &  S_{\mathrm{out},4}^{(n)} 
 = (\alpha^{4})^n
\int dt_{n} d\mbf{r}_{n} \cdots \int dt_{1} d\mbf{r}_{1}
 \int d\mbf{r}_{n''} \cdots d\mbf{r}_{1''} 
\int d\mbf{r}_{n'} \cdots d\mbf{r}_{1'}  \times \nonumber\\&
\int_{}^{} d\bar{t}_{n}    d\bar{\tau}_{n} 
F_{1,n}(\mbf{r}_0,\bar{t}_{n}^{}) 
F_{2,n}(\mbf{r}_0,\tau_{n}^{}) \cdots
\int_{}^{} d\bar{t}_{1} d\bar{\tau}_{1}  F_{1,1}(\mbf{r}_0,\bar{t}_{1}^{})
F_{2,1}(\mbf{r}_0,\tau_{1}^{}) 
\exp({
-i \tilde{\omega}_{\mbf{k}_{1}}^{} \bar{\tau}_{1}^{}
\cdots 
-i \tilde{\omega}_{\mbf{k}_{n}}^{} \bar{\tau}_{n}^{}
} )
\nonumber\\&
\sum_{\substack{\{\mu_{s_{j}}, \mu_{p_{j}}\}\\
\{\mbf{k}_{s_{j}}, \mbf{k}_{p_{j}}\} }}
\tilde{c}_{}^{2n} 
\Bigg(\frac{\lvert \varepsilon_{{\mbf{k}_{s_{1}}},\mu_{s_{1}}}^{*}\cdot  \varepsilon_{{\mbf{k}_{p_{1}}},\mu_{p_{1}}}^{} \rvert  }{\sqrt{\omega_{\mbf{k}_{s_{1}}} \omega_{\mbf{k}_{p_{1}}}}}  \cdots
\frac{\lvert \varepsilon_{{\mbf{k}_{s_{n}}},\mu_{s_{n}}}^{*}\cdot  \varepsilon_{{\mbf{k}_{p_{n}}},\mu_{p_{n}}}^{} \rvert  }{\sqrt{\omega_{\mbf{k}_{s_{n}}} \omega_{\mbf{k}_{p_{n}}}}} \Bigg)^2
\Big\{\sum_{\{m\}} \prod_{d,j',j''}
\delta_{\mbf{k}_{s_{j'}}\mbf{k}_{d}}\delta_{\mu_{s_{j'}}\mu_{d}} 
\delta_{\mbf{k}_{s_{j''}}\mbf{k}_{d}}\delta_{\mu_{s_{j''}}\mu_{d}} \Big\} \nonumber\\&
\quad \Big\{ K_{0}^{(1)}(\mbf{r}_{1'}^{};\mbf{r}_{1''}^{};\bar{\tau}_{1},\bar{t}_{1}) \cdots K_{0}^{(1)}(\mbf{r}_{n'}^{};\mbf{r}_{n''}^{};\bar{\tau}_{n},\bar{t}_{n}) \Big\} 
\exp({-i \tilde{\mbf{q}}_{1'}^{} \cdot\mbf{r}_{1'}^{}
\cdots 
-i \tilde{\mbf{q}}_{n'}^{} \cdot\mbf{r}_{n'}^{}
+i \tilde{\mbf{q}}_{n''}^{} \cdot\mbf{r}_{n''}^{}
\cdots 
+i \tilde{\mbf{q}}_{1''}^{} \cdot\mbf{r}_{1''}^{} )}.
\end{align} 
\end{widetext}
Here, $F_{1,j}(\mbf{r}_0,\bar{t}_{j}^{})=|\mathcal{E}_0^{}(\mbf{r}_{0},\bar{t}_{j} )|^2$ denotes the intensity of the incoming field and $F_{2,j}(\mbf{r}_0,\bar{\tau}_{j}^{})=\exp{(-\Gamma_{0,j} \bar{\tau}_j^2})$, where $\Gamma_{0,j} =2 \ln{2}/ \tau_{0,j}^2$, with $\tau_{0,j}$ being the temporal width of the pulse.
A more constrained approximation can be invoked at this point by factorizing the two-point density correlation functions in terms of products of means of electronic density operators.
It allows one to rewrite the electronic density correlation function appearing in the last line in Eq.~(\ref{eqn:sigbarefullfac3}) in the form
\begin{align}
&   K_{0}^{(1)}(\mbf{r}_{1'}^{};\mbf{r}_{1''}^{};\bar{\tau}_{1},\bar{t}_{1}) \cdots K_{0}^{(1)}(\mbf{r}_{n'}^{};\mbf{r}_{n''}^{};\bar{\tau}_{n},\bar{t}_{n}) 
\nonumber\\&
 \approx \ex{\mbf{n}(\mbf{r}_{1'}^{},\bar{\tau}_{1},\bar{t}_{1} )} 
\ex{\mbf{n}(\mbf{r}_{1''}^{},\bar{\tau}_{1},\bar{t}_{1})} 
\nonumber\\& 
 \qquad \cdots\ex{\mbf{n}(\mbf{r}_{n'}^{},\bar{\tau}_{n},\bar{t}_{n})}  
\ex{\mbf{n}(\mbf{r}_{n''}^{},\bar{\tau}_{n},\bar{t}_{n})}.
\end{align}
Additionally, we assume that the scattered modes are assigned to a particular set of predesignated detection modes. It amounts to 
neglecting the combinatorial nature of the corresponding correlation function. Moreover, if the mean of electronic density operators does not depend significantly on the temporal delay variables $\bar{\tau}_{j}^{}$, we may drop the argument in the expression and proceed to carry out the corresponding Fourier integral. If the detected frequencies are assumed to lie within the vicinity of the incoming frequencies, i.e., $\omega_{\mbf{k}_{s}} \approx \omega_{\mbf{k}_{p}}$, the Fourier integral of $F_{2,j}(\mbf{r}_0,\bar{\tau}_{j}^{})$ contains a frequency domain Gaussian function that is finite over a negligibly narrow range of frequencies. The exponent in the latter can be replaced by unity, i.e.,
$ \sqrt{\pi/\Gamma_{0,j}} \exp(-\tilde{\omega}_{\mbf{k}_j}^2/\Gamma_{0,j})
\approx \sqrt{\pi/\Gamma_{0,j}}$. 
Using these approximations, the expression in Eq.~(\ref{eqn:sigbarefullfac3}) can be symmetrized. For an approximation that drops time-dependence of the means of density operators entirely, the two temporal integrals decouple and can be carried out separately. These hierarchical sets of approximations are presented in the form
\begin{widetext}
\begin{subequations}
\begin{align}\label{eqn:sigbarefullfac4}
    & S_{\mathrm{out}, 5}^{(n)} =
 \prod_{j=1}^{n} \tilde{c}_{}^{2} \sqrt{\frac{\pi}{\Gamma_{0,j}}}
\int_{}^{}  d\bar{t}_{j}   F_{1,j}(\mbf{r}_0,\bar{t}_{j}^{})  
%
\times \alpha^{4}
\sum_{\substack{\mu_{s_j}, \mu_{p_j}\\
\mbf{k}_{s_j}, \mbf{k}_{p_j}}} 
\frac{\lvert \varepsilon_{{\mbf{k}_{s_{j}}},\mu_{s_{j}}}^{*}\cdot  \varepsilon_{{\mbf{k}_{p_{j}}},\mu_{p_{j}}}^{} \rvert^2  }{\omega_{\mbf{k}_{s_{j}}} \omega_{\mbf{k}_{p_{j}}}} \times
\Big|\int_{}^{} d\mbf{r}_{j}
e^{i \tilde{\mbf{q}}_{j}^{} \cdot \mbf{r}_{j}^{} }\ex{\mbf{n}(\mbf{r}_{j}^{},\bar{t}_{j}^{})} \Big|^2 \\
& \qquad\approx  \prod_{j=1}^{n} \Big\{\tilde{c}_{}^{2} \sqrt{\frac{\pi}{\Gamma_{0,j}}}
\int_{}^{}  d\bar{t}_{j}   F_{1,j}(\mbf{r}_0,\bar{t}_{j}^{})  \Big\}
%
\times \alpha^{4}
\sum_{\substack{\mu_{s_j}, \mu_{p_j}\\
\mbf{k}_{s_j}, \mbf{k}_{p_j}}} 
\frac{\lvert \varepsilon_{{\mbf{k}_{s_{j}}},\mu_{s_{j}}}^{*}\cdot  \varepsilon_{{\mbf{k}_{p_{j}}},\mu_{p_{j}}}^{} \rvert^2  }{\omega_{\mbf{k}_{s_{j}}} \omega_{\mbf{k}_{p_{j}}}} \times
\Big|\int_{}^{} d\mbf{r}_{j}
e^{i \tilde{\mbf{q}}_{j}^{} \cdot \mbf{r}_{j}^{} }\ex{\mbf{n}(\mbf{r}_{j}^{})} \Big|^2. \label{eqn:approxclassic}
\end{align}
\end{subequations}
\end{widetext}
Eq.~(\ref{eqn:approxclassic}), for the case of $n=1$, yields the diffraction signal which frequently appears in the literature. That expression may be thought of as a limit recoverable following a set of approximations in which dressing of electronic densities by classical-driving has been assumed. Traditionally, this limiting case, for the lowest order, was derived by considering the first-order perturbation in the photon-matter interaction under the Born approximation. Such considerations are solely based on the arguments regarding the field-matter interaction strength. The Born-approximation for the multiphoton case would imply the mutual independence of the scattering events. The presumptive application of such phenomenology leads to the factorizability of the scattering cross section in terms of mutually independent contributions. Within our treatment, this limit is recovered via a more rigorous and systematic procedure. In passing, we note that a decomposition of the average electron density in terms of contributions from atomic centers located at $\mbf{R}_p$ can be used to re-express the density scattering factors in terms of a linear combination of contributions at positions $p$ arising from atomic density scattering factors.
Such exercises have been taken up in several articles previously, notably in \cite{santra2008concepts, bennett2014time}. Following algebraic simplifications, it was concluded that the contributions to the diffractive imaging can be decomposed in terms of one- and two-particle contributions. In the general framework adopted in this article, the scattering target is treated as a single object, even it consists of several constituents e.g., many molecules of the same species. Therefore, we have refrained from adopting such a decomposition in terms of one- and two-particle contributions.
\section{Conclusions and outlook}
In this article, we formulated a theory of multiphoton x-ray diffraction, with a focus on the role of higher-order photonic and electronic correlations. The correlations and fluctuations associated with the outgoing photonic modes were shown to contain useful and sometimes crucial information for the electronic matter correlations. We also pointed out the physical origins of the correlation resources that may be exploited to design controlled diffractive imaging measurements. For specific physical situations, these correlation-generating resources may act in unison to give rise to novel signal components that contain a wealth of information regarding the underlying material structure.
These measurements may become crucial for designing advanced structural probes for complex correlated matter \cite{abbamonte2012resonant, gan2013crystallographic, cavalleri2006tracking, huang2021electron}, e.g., quantum materials with novel topological properties \cite{tabis2017synchrotron}, strange metals \cite{hartnoll2022colloquium, mousatov2020theory, thornton2022jamming} and quantum materials and molecules in engineered electromagnetic environments \cite{debnath2020entangled, 10.3389/fphy.2022.879113}. In general, the spatially resolved dielectric function of the corresponding material at higher energies can be probed and the correlation signatures, should they appear close to the ground state, can be investigated. It may provide a complementary technique to the widely available photoemission techniques for probing long-range electronic correlations.
Specifically, we emphasized the role of different approximations related to individual physical resources that lead to the factorization of the correlation functions.\\
The role of induced electronic current was not discussed extensively in this article. Diffractive scattering signals in the presence of the electronic current bear signatures of dynamical electronic resonances. Explicit considerations were made in omitting the relevant term, and neglecting diffractive scattering in presence of radiation damage. The physical processes which are sources of the radiation damage e.g., multiphoton ionization, electronic relaxation following ionization, relaxation mediated x-ray fluorescence, electronic rearrangement assisted Auger processes, electron-electron scattering etc. are not considered here. A diagrammatic extension to the presented theory, which systematically includes these roles, is the topic of upcoming work.
Additionally, explicit inclusion of the non-equilibrium structural dynamics and thermalization were omitted as well. This particular assumption is well justified in the situation where the scattering centers are far in space-time. In these situations, the scattering events at one instance does not exert dynamical influence on the other. The spatio-temporal coherence properties of the pulse are the key controlling factor that allows these features to be revealed.
Additionally, we have not presented any explicit treatment of the retardation effects, electromagnetic vacuum-induced corrections to the signal, and multiple scattering effects. Dispersive corrections are often less significant away from resonance in the elastic scattering regime. However, they may have considerable influence in altering the intrinsic coherence length scales in certain materials.
We also omitted the final state effects which may arise from the interaction of the scattered photons with the residual electronic polarization in the sample. This has been done by using vacuum averaging of the detected-scattered field mode correlation function. Any generalized detection process involving time-frequency, position-momentum filtered monitoring of the scattered modes can be easily incorporated by extending the averaging procedure. The joint space-time-frequency-momentum gating is typically introduced via a modulating function \cite{dorfman2012nonlinear, dorfman2016time, bennett2014time}. It can be systematically introduced by replacing the outgoing fields with their gated counterparts and evaluating the field contractions. Consequently, the detector response function is reexpressed in a generalized form prior to performing the detector convolution. The observed signal, following such convolution, depends on the resolution properties of the detector, as indicated by the properties of the gating functions.\\
However, we note that the incorporation of dynamical pathways related to the above-mentioned features may restrict the application of the factorization hierarchy. In practice, distinct signatures that point towards the existence of factorizable limits may need the full suite of correlation analyses. Simulation of the multipoint electronic correlation functions is a challenging task \cite{bonitz2020ab, dornheim2020nonlinear, dornheim2021nonlinear, dornheim2022analyzing}. Rate equations involving time-dependent electronic configurations have shown early promise \cite{son2011impact, jurek2016xmdyn, rohringer2006configuration}, which may provide a multiscale simulation toolbox towards the goal in the specified parameter regime. An extension of the rate-equation framework incorporating the inter-configuration coherence is underway. Such an algorithm, still numerically expensive, may provide a realistic way to investigate the factorization hierarchy. This, in turn, would require unique data-inversion algorithms.\\
In recent years, theoretical proposals concerning the usage of novel multimode radiation sources \cite{asban2019quantum} as investigative tools in structural biology have been made. The presented integral equation based formulation can easily accommodate them via the basic modification of the incoming field correlation functions. Moreover, optimally tailored incoming photonic field properties that selectively amplify specific parts of the signal can be found by employing control algorithms \cite{rothman2005observable, rothman2005quantum, rothman2006exploring, beltrani2009exploring, dominy2012dynamic}.\\
The inaccessibility of the phase in the intensity-detected scattering measurements has been a persistent problem that prevents the unambiguous inversion of the signal. It has been addressed in the case of crystalline and non-crystalline finite-sized samples via oversampling and iterative algorithms \cite{miao2000oversampling, miao2000possible}. In a set of forward-looking proposals, which may become challenging to implement at the current state-of-the-art, Mukamel et al. and others have proposed to resolve the problem with the help of phase-sensitive coincidence-detected diffractive scattering measurements \cite{asban2019quantum, ye2019imaging}. This proposal holds significant promise if it can be ingeniously combined with advanced detector technologies. The theoretical development in this paper requires minimal modification to incorporate and extend the proposed approach.\\
One of the consequences of monitoring fewer detection modes was that the available reconstruction algorithms assume the instantaneous electronic densities to be the central quantities for the posterior data inversion. It was shown that the multipoint density correlation functions are the generalized quantities carrying information about the distribution of the scattering centers. The presented paper lays the groundwork for the identification of specific terms in the matter correlation functions related to the relevant features in the diffraction signals.
The analysis, possibly, sets the ground for more accurate interpretations of the snapshot diffraction measurements, and reconstruction algorithms beyond density-based heuristics \cite{hosseinizadeh2021few, moths2011bayesian, cruz2021selecting, fancher2016use, dashti2020retrieving, ourmazd2022structural, bobkov2015sorting} and provides a theoretical framework for further developments pertinent to the single-particle imaging using FEL radiation sources \cite{lorenz2012impact, vartanyants2007coherent}. One of the promising applications of such an elaborate description lies in time-domain structural biology. Work in this direction is already underway by the authors.
\appendix
\section{Note on the combinatorial factorization of correlation functions}
In this appendix, we present the combinatorial expressions given in the main text [i.e., Eqs.~(\ref{eqn:sigcorr}), (\ref{eqn:fieldcorrfac}), and (\ref{eqn:matcorrfac})] for readability. The abbreviated expressions for factorized correlation functions related to two and three-photon diffractive scattering are expanded here. These expressions can also be used to rationalize the pairings of vertices for the diagrams in Figs.~\ref{fig:supdiag12} and \ref{fig:supdiag03}.\\
The correlation function that indicates the combinatorial detection pathways, for $n=2$, can be expressed as
\begin{widetext}
\begin{align}\label{eqn:AppDs2}
   & D_{s}^{(2)}(\mbf{r}_1,t_1,\mbf{r}_2, t_2;\mbf{r}_{1'}^{},t_{1'}^{},\mbf{r}_{2'}^{},t_{2'}^{};\mbf{r}_{2''}^{},t_{2''}^{},\mbf{r}_{1''}^{},t_{1''}^{})
\nonumber\\&    = 
    D_{s}^{(1)}(\mbf{r}_1,t_1; \mbf{r}_{1'}^{}, t_{1'}^{}) D_{s}^{(1),*}(\mbf{r}_1,t_1; \mbf{r}_{1''}^{},t_{1''}^{}) 
    \times 
    D_{s}^{(1)}(\mbf{r}_2,t_2; \mbf{r}_{2'}^{}, t_{2'}^{}) D_{s}^{(1),*}(\mbf{r}_2,t_2; \mbf{r}_{2''}^{}, t_{2''}^{}) \nonumber\\&
   +     D_{s}^{(1)}(\mbf{r}_1,t_1; \mbf{r}_{1'}^{}, t_{1'}^{}) D_{s}^{(1),*}(\mbf{r}_1,t_1; \mbf{r}_{2''}^{},t_{2''}^{}) 
   \times  
   D_{s}^{(1)}(\mbf{r}_2,t_2; \mbf{r}_{2'}^{}, t_{2'}^{}) D_{s}^{(1),*}(\mbf{r}_2,t_2; \mbf{r}_{1''}^{}, t_{1''}^{}) \nonumber\\
   &+     D_{s}^{(1)}(\mbf{r}_1,t_1; \mbf{r}_{2'}^{}, t_{2'}^{}) D_{s}^{(1),*}(\mbf{r}_1,t_1; \mbf{r}_{1''}^{},t_{1''}^{}) 
   \times  
   D_{s}^{(1)}(\mbf{r}_2,t_2; \mbf{r}_{1'}^{}, t_{1'}^{}) D_{s}^{(1),*}(\mbf{r}_2,t_2; \mbf{r}_{2''}^{}, t_{2''}^{}) \nonumber\\
   &+     D_{s}^{(1)}(\mbf{r}_1,t_1; \mbf{r}_{2'}^{}, t_{2'}^{}) D_{s}^{(1),*}(\mbf{r}_1,t_1; \mbf{r}_{2''}^{},t_{2''}^{}) 
   \times  
   D_{s}^{(1)}(\mbf{r}_2,t_2; \mbf{r}_{1'}^{}, t_{1'}^{}) D_{s}^{(1),*}(\mbf{r}_2,t_2; \mbf{r}_{1''}^{}, t_{1''}^{}) .
\end{align}
\end{widetext}
In the above, we note that the scattered modes can be paired with the detection modes in four (i.e., $m=4$) distinct ways.
Similarly, for the case of $n=3$ we have $m=36$ terms which can be expressed as, 
\begin{widetext}
\begin{align}\label{eqn:AppDs3}
&D_{s}^{(3)}(\mbf{r}_1,t_1,\mbf{r}_1, t_1,\mbf{r}_2, t_2,\mbf{r}_3, t_3;\mbf{r}_{1'}^{},t_{1'}^{},\mbf{r}_{2'}^{},t_{2'}^{},\mbf{r}_{3'}^{},t_{3'}^{} ;\mbf{r}_{3''}^{},t_{3''}^{},\mbf{r}_{2''}^{},t_{2''}^{},\mbf{r}_{1''}^{},t_{1''}^{})\nonumber\\
    &= \Big\{
    D_{s}^{(1)}(\mbf{r}_{1}^{},t_{1}^{} ;\mbf{r}_{1'}^{},t_{1'}^{}) D_{s}^{(1)}(\mbf{r}_{2}^{},t_{2}^{} ;\mbf{r}_{2'}^{},t_{2'}^{}) 
   D_{s}^{(1)}(\mbf{r}_{3}^{},t_{3}^{} ;\mbf{r}_{3'}^{},t_{3'}^{})
   \nonumber\\
   & +  D_{s}^{(1)}(\mbf{r}_{1}^{},t_{1}^{} ;\mbf{r}_{2'}^{},t_{2'}^{})  D_{s}^{(1)}(\mbf{r}_{2}^{},t_{2}^{} ;\mbf{r}_{1'}^{},t_{1'}^{}) 
   D_{s}^{(1)}(\mbf{r}_{3}^{},t_{3}^{} ;\mbf{r}_{3'}^{},t_{3'}^{})
   \nonumber\\ 
   &+ D_{s}^{(1)}(\mbf{r}_{1}^{},t_{1}^{} ;\mbf{r}_{3'}^{},t_{3'}^{}) 
   D_{s}^{(1)}(\mbf{r}_{2}^{},t_{2}^{} ;\mbf{r}_{2'}^{},t_{2'}^{}) 
   D_{s}^{(1)}(\mbf{r}_{3}^{},t_{3}^{} ;\mbf{r}_{1'}^{},t_{1'}^{})
   \nonumber\\
 & +D_{s}^{(1)}(\mbf{r}_{1}^{},t_{1}^{} ;\mbf{r}_{1'}^{},t_{1'}^{}) 
   D_{s}^{(1)}(\mbf{r}_{2}^{},t_{2}^{} ;\mbf{r}_{3'}^{},t_{3'}^{}) 
   D_{s}^{(1)}(\mbf{r}_{3}^{},t_{3}^{} ;\mbf{r}_{2'}^{},t_{2'}^{})
   \nonumber\\
  & +D_{s}^{(1)}(\mbf{r}_{1}^{},t_{1}^{} ;\mbf{r}_{2'}^{},t_{2'}^{}) 
   D_{s}^{(1)}(\mbf{r}_{2}^{},t_{2}^{} ;\mbf{r}_{3'}^{},t_{3'}^{}) 
   D_{s}^{(1)}(\mbf{r}_{3}^{},t_{3}^{} ;\mbf{r}_{1'}^{},t_{1'}^{})
   \nonumber\\
  &+ D_{s}^{(1)}(\mbf{r}_{1}^{},t_{1}^{} ;\mbf{r}_{3'}^{},t_{3'}^{}) 
   D_{s}^{(1)}(\mbf{r}_{2}^{},t_{2}^{} ;\mbf{r}_{1'}^{},t_{1'}^{}) 
    D_{s}^{(1)}(\mbf{r}_{3}^{},t_{3}^{} ;\mbf{r}_{2'}^{},t_{2'}^{})\Big\} \nonumber\\&
 \times\Big\{
    D_{s}^{(1)*}(\mbf{r}_{1}^{},t_{1}^{} ;\mbf{r}_{1''}^{},t_{1''}^{}) D_{s}^{(1)*}(\mbf{r}_{2}^{},t_{2}^{} ;\mbf{r}_{2''}^{},t_{2''}^{}) 
   D_{s}^{(1)*}(\mbf{r}_{3}^{},t_{3}^{} ;\mbf{r}_{3''}^{},t_{3''}^{})
   \nonumber\\
   & +  D_{s}^{(1)*}(\mbf{r}_{1}^{},t_{1}^{} ;\mbf{r}_{2''}^{},t_{2''}^{})  D_{s}^{(1)*}(\mbf{r}_{2}^{},t_{2}^{} ;\mbf{r}_{1''}^{},t_{1''}^{}) 
   D_{s}^{(1)*}(\mbf{r}_{3}^{},t_{3}^{} ;\mbf{r}_{3''}^{},t_{3''}^{})
   \nonumber\\ 
   &+ D_{s}^{(1)*}(\mbf{r}_{1}^{},t_{1}^{} ;\mbf{r}_{3''}^{},t_{3''}^{}) 
   D_{s}^{(1)*}(\mbf{r}_{2}^{},t_{2}^{} ;\mbf{r}_{2''}^{},t_{2''}^{}) 
   D_{s}^{(1)*}(\mbf{r}_{3}^{},t_{3}^{} ;\mbf{r}_{1''}^{},t_{1''}^{})
   \nonumber\\
 & +D_{s}^{(1)*}(\mbf{r}_{1}^{},t_{1}^{} ;\mbf{r}_{1''}^{},t_{1''}^{}) 
   D_{s}^{(1)*}(\mbf{r}_{2}^{},t_{2}^{} ;\mbf{r}_{3''}^{},t_{3''}^{}) 
   D_{s}^{(1)*}(\mbf{r}_{3}^{},t_{3}^{} ;\mbf{r}_{2''}^{},t_{2''}^{})
   \nonumber\\
  & +D_{s}^{(1)*}(\mbf{r}_{1}^{},t_{1}^{} ;\mbf{r}_{2''}^{},t_{2''}^{}) 
   D_{s}^{(1)*}(\mbf{r}_{2}^{},t_{2}^{} ;\mbf{r}_{3''}^{},t_{3''}^{}) 
   D_{s}^{(1)*}(\mbf{r}_{3}^{},t_{3}^{} ;\mbf{r}_{1''}^{},t_{1''}^{})
   \nonumber\\
  &+ D_{s}^{(1)*}(\mbf{r}_{1}^{},t_{1}^{} ;\mbf{r}_{3''}^{},t_{3''}^{}) 
   D_{s}^{(1)*}(\mbf{r}_{2}^{},t_{2}^{} ;\mbf{r}_{1''}^{},t_{1''}^{}) 
    D_{s}^{(1)*}(\mbf{r}_{3}^{},t_{3}^{} ;\mbf{r}_{2''}^{},t_{2''}^{})\Big\}    
\end{align}
\end{widetext}
Following an analogous exercise, the factorization of the incoming field correlation functions can be performed as well. In doing so, only the $2$-tuples arising from the pairings across branches were kept. This particular approximation neglects the effect of extended spatio-temporal field coherence on densities.
For the two-point incoming field correlation function it yields two terms (i.e., $w=2$) given by
\begin{align}\label{eqn:AppDp2}
  &  \tilde{D}_{p}^{(2)}(\mbf{r}_{1'}^{},t_{1'}^{},\mbf{r}_{2'}^{},t_{2'}^{};\mbf{r}_{2''}^{},t_{2''}^{},\mbf{r}_{1''}^{},t_{1''}^{})\nonumber\\&
   =  \tilde{D}_{p}^{(1)}(\mbf{r}_{1'}^{},t_{1'}^{} ;\mbf{r}_{1''}^{},t_{1''}^{}) \times 
   \tilde{D}_{p}^{(1)}(\mbf{r}_{2'}^{},t_{2'}^{} ;\mbf{r}_{2''}^{},t_{2''}^{}) \nonumber\\&
    +  \tilde{D}_{p}^{(1)}(\mbf{r}_{1'}^{},t_{1'}^{} ;\mbf{r}_{2''}^{},t_{2''}^{}) \times \tilde{D}_{p}^{(1)}(\mbf{r}_{2'}^{},t_{2'}^{} ;\mbf{r}_{1''}^{},t_{1''}^{}) ,
\end{align}
A similar exercise for the three-point incoming field correlation function yields six terms (i.e., $w=6$). In doing so, we have omitted any intermediate correlation functions of next higher-order, i.e., four-point ones. The expressions are given by
\begin{widetext}
\begin{align}\label{eqn:AppDp3}
   & \tilde{D}_{p}^{(3)}(\mbf{r}_{1'}^{},t_{1'}^{},\mbf{r}_{2'}^{},t_{2'}^{},\mbf{r}_{3'}^{},t_{3'}^{} ;\mbf{r}_{3''}^{},t_{3''}^{},\mbf{r}_{2''}^{},t_{2''}^{},\mbf{r}_{1''}^{},t_{1''}^{})
   \nonumber\\&
   =  \tilde{D}_{p}^{(1)}(\mbf{r}_{1'}^{},t_{1'}^{} ;\mbf{r}_{1''}^{},t_{1''}^{}) \tilde{D}_{p}^{(1)}(\mbf{r}_{2'}^{},t_{2'}^{} ;\mbf{r}_{2''}^{},t_{2''}^{}) 
   \tilde{D}_{p}^{(1)}(\mbf{r}_{3'}^{},t_{3'}^{} ;\mbf{r}_{3''}^{},t_{3''}^{})
   \nonumber\\
   & +  \tilde{D}_{p}^{(1)}(\mbf{r}_{1'}^{},t_{1'}^{} ;\mbf{r}_{2''}^{},t_{2''}^{})  \tilde{D}_{p}^{(1)}(\mbf{r}_{2'}^{},t_{2'}^{} ;\mbf{r}_{1''}^{},t_{1''}^{}) 
   \tilde{D}_{p}^{(1)}(\mbf{r}_{3'}^{},t_{3'}^{} ;\mbf{r}_{3''}^{},t_{3''}^{})
   \nonumber\\ 
   &+ \tilde{D}_{p}^{(1)}(\mbf{r}_{1'}^{},t_{1'}^{} ;\mbf{r}_{3''}^{},t_{3''}^{}) 
   \tilde{D}_{p}^{(1)}(\mbf{r}_{2'}^{},t_{2'}^{} ;\mbf{r}_{2''}^{},t_{2''}^{}) 
   \tilde{D}_{p}^{(1)}(\mbf{r}_{3'}^{},t_{3'}^{} ;\mbf{r}_{1''}^{},t_{1''}^{})
   \nonumber\\
 & +\tilde{D}_{p}^{(1)}(\mbf{r}_{1'}^{},t_{1'}^{} ;\mbf{r}_{1''}^{},t_{1''}^{}) 
   \tilde{D}_{p}^{(1)}(\mbf{r}_{2'}^{},t_{2'}^{} ;\mbf{r}_{3''}^{},t_{3''}^{}) 
   \tilde{D}_{p}^{(1)}(\mbf{r}_{3'}^{},t_{3'}^{} ;\mbf{r}_{2''}^{},t_{2''}^{})
   \nonumber\\
  & +\tilde{D}_{p}^{(1)}(\mbf{r}_{1'}^{},t_{1'}^{} ;\mbf{r}_{2''}^{},t_{2''}^{}) 
   \tilde{D}_{p}^{(1)}(\mbf{r}_{2'}^{},t_{2'}^{} ;\mbf{r}_{3''}^{},t_{3''}^{}) 
   \tilde{D}_{p}^{(1)}(\mbf{r}_{3'}^{},t_{3'}^{} ;\mbf{r}_{1''}^{},t_{1''}^{})
   \nonumber\\
  &+ \tilde{D}_{p}^{(1)}(\mbf{r}_{1'}^{},t_{1'}^{} ;\mbf{r}_{3''}^{},t_{3''}^{}) 
   \tilde{D}_{p}^{(1)}(\mbf{r}_{2'}^{},t_{2'}^{} ;\mbf{r}_{1''}^{},t_{1''}^{}) 
    \tilde{D}_{p}^{(1)}(\mbf{r}_{3'}^{},t_{3'}^{} ;\mbf{r}_{2''}^{},t_{2''}^{}).
\end{align}
\end{widetext}
The assumption that the incoming field properties dominantly dictate the factorization of the electronic density correlation functions, leads to a factorization rule that is identical to the previous pairings. Consequently, the number of combinatorial terms being generated is also the same.
For $n=2$, four-point density correlation functions can be factorized as follows
\begin{align}\label{eqn:AppK2}
    & K_0^{(2)}(\mbf{r}_{1'}^{},t_{1'}^{},\mbf{r}_{2'}^{},t_{2'}^{} ;\mbf{r}_{2''}^{},t_{2''}^{},\mbf{r}_{1''}^{},t_{1''}^{}) 
\nonumber\\
   &=  K_{0}^{(1)}(\mbf{r}_{1'}^{},t_{1'}^{} ;\mbf{r}_{1''}^{},t_{1''}^{})  
   K_{0}^{(1)}(\mbf{r}_{2'}^{},t_{2'}^{} ;\mbf{r}_{2''}^{},t_{2''}^{}) \nonumber\\&
    +  K_{0}^{(1)}(\mbf{r}_{1'}^{},t_{1'}^{} ;\mbf{r}_{2''}^{},t_{2''}^{})  K_{0}^{(1)}(\mbf{r}_{2'}^{},t_{2'}^{} ;\mbf{r}_{1''}^{},t_{1''}^{}).
\end{align}
Here, the four-point density correlation functions yielded two distinct terms. The neglected terms can be given as
\begin{align}\label{eqn:AppKaux2}
   & K_{0,\mathrm{aux}}^{(2)}(\mbf{r}_{1'}^{},t_{1'}^{},\mbf{r}_{2'}^{},t_{2'}^{} ;\mbf{r}_{2''}^{},t_{2''}^{},\mbf{r}_{1''}^{},t_{1''}^{}) \nonumber\\&
    =  K_{0}^{(1)}(\mbf{r}_{1'}^{},t_{1'}^{}
    ;\mbf{r}_{1''}^{},t_{1''}^{})  
    K_{0}^{(1)}(\mbf{r}_{2'}^{},t_{2'}^{} ;\mbf{r}_{2''}^{},t_{2''}^{}) 
\end{align}
Putting these two sets together, it yields $g=3$ terms. A more involved situation is offered by the case, $n=3$ which yields the following set containing six terms, as expected. They are given by
\begin{widetext}
\begin{align}\label{eqn:AppK3}
& K_0^{(3)}(\mbf{r}_{1'}^{},t_{1'}^{},\mbf{r}_{2'}^{},t_{2'}^{},\mbf{r}_{3'}^{},t_{3'}^{} ;\mbf{r}_{3''}^{},t_{3''}^{},\mbf{r}_{2''}^{},t_{2''}^{},\mbf{r}_{1''}^{},t_{1''}^{})  \nonumber\\
&=
K_{0}^{(1)}(\mbf{r}_{1'}^{},t_{1'}^{} ;\mbf{r}_{1''}^{},t_{1''}^{})
K_{0}^{(1)}(\mbf{r}_{2'}^{},t_{2'}^{} ;\mbf{r}_{2''}^{},t_{2''}^{})  K_{0}^{(1)}(\mbf{r}_{3'}^{},t_{3'}^{} ;\mbf{r}_{3''}^{},t_{3''}^{})
\nonumber\\&
+ K_{0}^{(1)}(\mbf{r}_{1'}^{},t_{1'}^{} ;\mbf{r}_{2''}^{},t_{2''}^{}) K_{0}^{(1)}(\mbf{r}_{2'}^{},t_{2'}^{} ;\mbf{r}_{1''}^{},t_{1''}^{}) K_{0}^{(1)}(\mbf{r}_{3'}^{},t_{3'}^{} ;\mbf{r}_{3''}^{},t_{3''}^{})
\nonumber\\&
+ K_{0}^{(1)}(\mbf{r}_{1'}^{},t_{1'}^{} ;\mbf{r}_{3''}^{},t_{3''}^{}) K_{0}^{(1)}(\mbf{r}_{2'}^{},t_{2'}^{} ;\mbf{r}_{2''}^{},t_{2''}^{}) 
K_{0}^{(1)}(\mbf{r}_{3'}^{},t_{3'}^{} ;\mbf{r}_{1''}^{},t_{1''}^{})
\nonumber\\&
+K_{0}^{(1)}(\mbf{r}_{1'}^{},t_{1'}^{} ;\mbf{r}_{1''}^{},t_{1''}^{})  K_{0}^{(1)}(\mbf{r}_{2'}^{},t_{2'}^{} ;\mbf{r}_{3''}^{},t_{3''}^{})  K_{0}^{(1)}(\mbf{r}_{3'}^{},t_{3'}^{} ;\mbf{r}_{2''}^{},t_{2''}^{})
\nonumber\\&
+K_{0}^{(1)}(\mbf{r}_{1'}^{},t_{1'}^{} ;\mbf{r}_{2''}^{},t_{2''}^{}) K_{0}^{(1)}(\mbf{r}_{2'}^{},t_{2'}^{} ;\mbf{r}_{3''}^{},t_{3''}^{}) 
K_{0}^{(1)}(\mbf{r}_{3'}^{},t_{3'}^{} ;\mbf{r}_{1''}^{},t_{1''}^{})
\nonumber\\&
+ K_{0}^{(1)}(\mbf{r}_{1'}^{},t_{1'}^{} ;\mbf{r}_{3''}^{},t_{3''}^{}) K_{0}^{(1)}(\mbf{r}_{2'}^{},t_{2'}^{} ;\mbf{r}_{1''}^{},t_{1''}^{}) 
K_{0}^{(1)}(\mbf{r}_{3'}^{},t_{3'}^{} ;\mbf{r}_{2''}^{},t_{2''}^{}).
\end{align}
\end{widetext}
The neglected terms in this case are given by
\begin{widetext}
\begin{align}\label{eqn:AppKaux3}
& K_{0,\mathrm{aux}}^{(3)}(\mbf{r}_{1'}^{},t_{1'}^{},\mbf{r}_{2'}^{},t_{2'}^{},\mbf{r}_{3'}^{},t_{3'}^{} ;\mbf{r}_{3''}^{},t_{3''}^{},\mbf{r}_{2''}^{},t_{2''}^{},\mbf{r}_{1''}^{},t_{1''}^{}) = \nonumber\\
&
K_{0}^{(1)}(\mbf{r}_{1'}^{},t_{1'}^{} ;\mbf{r}_{1''}^{},t_{1''}^{})
K_{0}^{(1)}(\mbf{r}_{2'}^{},t_{2'}^{} ;\mbf{r}_{3'}^{},t_{3'}^{})  K_{0}^{(1)}(\mbf{r}_{2''}^{},t_{2''}^{} ;\mbf{r}_{3''}^{},t_{3''}^{}) \nonumber\\&
K_{0}^{(1)}(\mbf{r}_{2'}^{},t_{2'}^{} ;\mbf{r}_{2''}^{},t_{2''}^{})
K_{0}^{(1)}(\mbf{r}_{1'}^{},t_{1'}^{} ;\mbf{r}_{3'}^{},t_{3'}^{})  K_{0}^{(1)}(\mbf{r}_{1''}^{},t_{1''}^{} ;\mbf{r}_{3''}^{},t_{3''}^{}) \nonumber\\&
K_{0}^{(1)}(\mbf{r}_{3'}^{},t_{3'}^{} ;\mbf{r}_{3''}^{},t_{3''}^{})
K_{0}^{(1)}(\mbf{r}_{2'}^{},t_{2'}^{} ;\mbf{r}_{3'}^{},t_{3'}^{})  K_{0}^{(1)}(\mbf{r}_{2''}^{},t_{2''}^{} ;\mbf{r}_{3''}^{},t_{3''}^{}) \nonumber\\&
K_{0}^{(1)}(\mbf{r}_{1'}^{},t_{1'}^{} ;\mbf{r}_{2''}^{},t_{2''}^{})
K_{0}^{(1)}(\mbf{r}_{2'}^{},t_{2'}^{} ;\mbf{r}_{3'}^{},t_{3'}^{})  K_{0}^{(1)}(\mbf{r}_{1''}^{},t_{1''}^{} ;\mbf{r}_{3''}^{},t_{3''}^{}) \nonumber\\&
K_{0}^{(1)}(\mbf{r}_{1'}^{},t_{1'}^{} ;\mbf{r}_{3''}^{},t_{3''}^{})
K_{0}^{(1)}(\mbf{r}_{2'}^{},t_{2'}^{} ;\mbf{r}_{3'}^{},t_{3'}^{})  K_{0}^{(1)}(\mbf{r}_{1''}^{},t_{1''}^{} ;\mbf{r}_{2''}^{},t_{2''}^{}) \nonumber\\&
K_{0}^{(1)}(\mbf{r}_{2'}^{},t_{2'}^{} ;\mbf{r}_{1''}^{},t_{1''}^{})
K_{0}^{(1)}(\mbf{r}_{1'}^{},t_{1'}^{} ;\mbf{r}_{3'}^{},t_{3'}^{})  K_{0}^{(1)}(\mbf{r}_{2''}^{},t_{2''}^{} ;\mbf{r}_{3''}^{},t_{3''}^{}) \nonumber\\&
K_{0}^{(1)}(\mbf{r}_{2'}^{},t_{2'}^{} ;\mbf{r}_{3''}^{},t_{3''}^{})
K_{0}^{(1)}(\mbf{r}_{1'}^{},t_{1'}^{} ;\mbf{r}_{3'}^{},t_{3'}^{})  K_{0}^{(1)}(\mbf{r}_{2''}^{},t_{2''}^{} ;\mbf{r}_{1''}^{},t_{1''}^{}) \nonumber\\&
K_{0}^{(1)}(\mbf{r}_{2'}^{},t_{2'}^{} ;\mbf{r}_{3''}^{},t_{3''}^{})
K_{0}^{(1)}(\mbf{r}_{2'}^{},t_{2'}^{} ;\mbf{r}_{1'}^{},t_{1'}^{})  K_{0}^{(1)}(\mbf{r}_{2''}^{},t_{2''}^{} ;\mbf{r}_{1''}^{},t_{1''}^{}) \nonumber\\&
K_{0}^{(1)}(\mbf{r}_{3'}^{},t_{3'}^{} ;\mbf{r}_{2''}^{},t_{2''}^{})
K_{0}^{(1)}(\mbf{r}_{2'}^{},t_{2'}^{} ;\mbf{r}_{1'}^{},t_{1'}^{})  K_{0}^{(1)}(\mbf{r}_{1''}^{},t_{1''}^{} ;\mbf{r}_{3''}^{},t_{3''}^{}) \nonumber\\&
\end{align}
\end{widetext}
As similar to the previous case, terms belonging to these two sets can be collected together to yield a total number of $g=15$ combinatorial terms.
For each of these cases, the parameters defining the intrinsic spatio-temporal coherence length scale
along with the coherence properties of the incoming field is expected to provide the physical basis for the factorization.

\bibliography{Ref_DebnathSantra01}

\begin{thebibliography}{68}%
\makeatletter
\providecommand \@ifxundefined [1]{%
 \@ifx{#1\undefined}
}%
\providecommand \@ifnum [1]{%
 \ifnum #1\expandafter \@firstoftwo
 \else \expandafter \@secondoftwo
 \fi
}%
\providecommand \@ifx [1]{%
 \ifx #1\expandafter \@firstoftwo
 \else \expandafter \@secondoftwo
 \fi
}%
\providecommand \natexlab [1]{#1}%
\providecommand \enquote  [1]{``#1''}%
\providecommand \bibnamefont  [1]{#1}%
\providecommand \bibfnamefont [1]{#1}%
\providecommand \citenamefont [1]{#1}%
\providecommand \href@noop [0]{\@secondoftwo}%
\providecommand \href [0]{\begingroup \@sanitize@url \@href}%
\providecommand \@href[1]{\@@startlink{#1}\@@href}%
\providecommand \@@href[1]{\endgroup#1\@@endlink}%
\providecommand \@sanitize@url [0]{\catcode `\\12\catcode `\$12\catcode
  `\&12\catcode `\#12\catcode `\^12\catcode `\_12\catcode `\%12\relax}%
\providecommand \@@startlink[1]{}%
\providecommand \@@endlink[0]{}%
\providecommand \url  [0]{\begingroup\@sanitize@url \@url }%
\providecommand \@url [1]{\endgroup\@href {#1}{\urlprefix }}%
\providecommand \urlprefix  [0]{URL }%
\providecommand \Eprint [0]{\href }%
\providecommand \doibase [0]{https://doi.org/}%
\providecommand \selectlanguage [0]{\@gobble}%
\providecommand \bibinfo  [0]{\@secondoftwo}%
\providecommand \bibfield  [0]{\@secondoftwo}%
\providecommand \translation [1]{[#1]}%
\providecommand \BibitemOpen [0]{}%
\providecommand \bibitemStop [0]{}%
\providecommand \bibitemNoStop [0]{.\EOS\space}%
\providecommand \EOS [0]{\spacefactor3000\relax}%
\providecommand \BibitemShut  [1]{\csname bibitem#1\endcsname}%
\let\auto@bib@innerbib\@empty
\bibitem [{\citenamefont {Chapman}\ \emph {et~al.}(2006)\citenamefont
  {Chapman}, \citenamefont {Barty}, \citenamefont {Bogan}, \citenamefont
  {Boutet}, \citenamefont {Frank}, \citenamefont {Hau-Riege}, \citenamefont
  {Marchesini}, \citenamefont {Woods}, \citenamefont {Bajt}, \citenamefont
  {Benner} \emph {et~al.}}]{chapman2006femtosecond}%
  \BibitemOpen
  \bibfield  {author} {\bibinfo {author} {\bibfnamefont {H.~N.}\ \bibnamefont
  {Chapman}}, \bibinfo {author} {\bibfnamefont {A.}~\bibnamefont {Barty}},
  \bibinfo {author} {\bibfnamefont {M.~J.}\ \bibnamefont {Bogan}}, \bibinfo
  {author} {\bibfnamefont {S.}~\bibnamefont {Boutet}}, \bibinfo {author}
  {\bibfnamefont {M.}~\bibnamefont {Frank}}, \bibinfo {author} {\bibfnamefont
  {S.~P.}\ \bibnamefont {Hau-Riege}}, \bibinfo {author} {\bibfnamefont
  {S.}~\bibnamefont {Marchesini}}, \bibinfo {author} {\bibfnamefont {B.~W.}\
  \bibnamefont {Woods}}, \bibinfo {author} {\bibfnamefont {S.}~\bibnamefont
  {Bajt}}, \bibinfo {author} {\bibfnamefont {W.~H.}\ \bibnamefont {Benner}},
  \emph {et~al.},\ }\bibfield  {title} {\bibinfo {title} {Femtosecond
  diffractive imaging with a soft-x-ray free-electron laser},\ }\href@noop {}
  {\bibfield  {journal} {\bibinfo  {journal} {Nature Physics}\ }\textbf
  {\bibinfo {volume} {2}},\ \bibinfo {pages} {839} (\bibinfo {year}
  {2006})}\BibitemShut {NoStop}%
\bibitem [{\citenamefont {Gaffney}\ and\ \citenamefont
  {Chapman}(2007)}]{gaffney2007imaging}%
  \BibitemOpen
  \bibfield  {author} {\bibinfo {author} {\bibfnamefont {K.}~\bibnamefont
  {Gaffney}}\ and\ \bibinfo {author} {\bibfnamefont {H.~N.}\ \bibnamefont
  {Chapman}},\ }\bibfield  {title} {\bibinfo {title} {Imaging atomic structure
  and dynamics with ultrafast x-ray scattering},\ }\href@noop {} {\bibfield
  {journal} {\bibinfo  {journal} {Science}\ }\textbf {\bibinfo {volume}
  {316}},\ \bibinfo {pages} {1444} (\bibinfo {year} {2007})}\BibitemShut
  {NoStop}%
\bibitem [{\citenamefont {Young}\ \emph {et~al.}(2018)\citenamefont {Young},
  \citenamefont {Ueda}, \citenamefont {G{\"u}hr}, \citenamefont {Bucksbaum},
  \citenamefont {Simon}, \citenamefont {Mukamel}, \citenamefont {Rohringer},
  \citenamefont {Prince}, \citenamefont {Masciovecchio}, \citenamefont {Meyer}
  \emph {et~al.}}]{young2018roadmap}%
  \BibitemOpen
  \bibfield  {author} {\bibinfo {author} {\bibfnamefont {L.}~\bibnamefont
  {Young}}, \bibinfo {author} {\bibfnamefont {K.}~\bibnamefont {Ueda}},
  \bibinfo {author} {\bibfnamefont {M.}~\bibnamefont {G{\"u}hr}}, \bibinfo
  {author} {\bibfnamefont {P.~H.}\ \bibnamefont {Bucksbaum}}, \bibinfo {author}
  {\bibfnamefont {M.}~\bibnamefont {Simon}}, \bibinfo {author} {\bibfnamefont
  {S.}~\bibnamefont {Mukamel}}, \bibinfo {author} {\bibfnamefont
  {N.}~\bibnamefont {Rohringer}}, \bibinfo {author} {\bibfnamefont {K.~C.}\
  \bibnamefont {Prince}}, \bibinfo {author} {\bibfnamefont {C.}~\bibnamefont
  {Masciovecchio}}, \bibinfo {author} {\bibfnamefont {M.}~\bibnamefont
  {Meyer}}, \emph {et~al.},\ }\bibfield  {title} {\bibinfo {title} {Roadmap of
  ultrafast x-ray atomic and molecular physics},\ }\href@noop {} {\bibfield
  {journal} {\bibinfo  {journal} {Journal of Physics B: Atomic, Molecular and
  Optical Physics}\ }\textbf {\bibinfo {volume} {51}},\ \bibinfo {pages}
  {032003} (\bibinfo {year} {2018})}\BibitemShut {NoStop}%
\bibitem [{\citenamefont {Marchesini}\ \emph {et~al.}(2003)\citenamefont
  {Marchesini}, \citenamefont {Chapman}, \citenamefont {Hau-Riege},
  \citenamefont {London}, \citenamefont {Szoke}, \citenamefont {He},
  \citenamefont {Howells}, \citenamefont {Padmore}, \citenamefont {Rosen},
  \citenamefont {Spence} \emph {et~al.}}]{marchesini2003coherent}%
  \BibitemOpen
  \bibfield  {author} {\bibinfo {author} {\bibfnamefont {S.}~\bibnamefont
  {Marchesini}}, \bibinfo {author} {\bibfnamefont {H.}~\bibnamefont {Chapman}},
  \bibinfo {author} {\bibfnamefont {S.}~\bibnamefont {Hau-Riege}}, \bibinfo
  {author} {\bibfnamefont {R.}~\bibnamefont {London}}, \bibinfo {author}
  {\bibfnamefont {A.}~\bibnamefont {Szoke}}, \bibinfo {author} {\bibfnamefont
  {H.}~\bibnamefont {He}}, \bibinfo {author} {\bibfnamefont {M.}~\bibnamefont
  {Howells}}, \bibinfo {author} {\bibfnamefont {H.}~\bibnamefont {Padmore}},
  \bibinfo {author} {\bibfnamefont {R.}~\bibnamefont {Rosen}}, \bibinfo
  {author} {\bibfnamefont {J.}~\bibnamefont {Spence}}, \emph {et~al.},\
  }\bibfield  {title} {\bibinfo {title} {Coherent x-ray diffractive imaging:
  applications and limitations},\ }\href@noop {} {\bibfield  {journal}
  {\bibinfo  {journal} {Optics Express}\ }\textbf {\bibinfo {volume} {11}},\
  \bibinfo {pages} {2344} (\bibinfo {year} {2003})}\BibitemShut {NoStop}%
\bibitem [{\citenamefont {Kotani}\ and\ \citenamefont
  {Shin}(2001)}]{kotani2001resonant}%
  \BibitemOpen
  \bibfield  {author} {\bibinfo {author} {\bibfnamefont {A.}~\bibnamefont
  {Kotani}}\ and\ \bibinfo {author} {\bibfnamefont {S.}~\bibnamefont {Shin}},\
  }\bibfield  {title} {\bibinfo {title} {Resonant inelastic x-ray scattering
  spectra for electrons in solids},\ }\href@noop {} {\bibfield  {journal}
  {\bibinfo  {journal} {Reviews of Modern Physics}\ }\textbf {\bibinfo {volume}
  {73}},\ \bibinfo {pages} {203} (\bibinfo {year} {2001})}\BibitemShut
  {NoStop}%
\bibitem [{\citenamefont {Yoneda}\ \emph {et~al.}(2015)\citenamefont {Yoneda},
  \citenamefont {Inubushi}, \citenamefont {Nagamine}, \citenamefont {Michine},
  \citenamefont {Ohashi}, \citenamefont {Yumoto}, \citenamefont {Yamauchi},
  \citenamefont {Mimura}, \citenamefont {Kitamura}, \citenamefont {Katayama}
  \emph {et~al.}}]{yoneda2015atomic}%
  \BibitemOpen
  \bibfield  {author} {\bibinfo {author} {\bibfnamefont {H.}~\bibnamefont
  {Yoneda}}, \bibinfo {author} {\bibfnamefont {Y.}~\bibnamefont {Inubushi}},
  \bibinfo {author} {\bibfnamefont {K.}~\bibnamefont {Nagamine}}, \bibinfo
  {author} {\bibfnamefont {Y.}~\bibnamefont {Michine}}, \bibinfo {author}
  {\bibfnamefont {H.}~\bibnamefont {Ohashi}}, \bibinfo {author} {\bibfnamefont
  {H.}~\bibnamefont {Yumoto}}, \bibinfo {author} {\bibfnamefont
  {K.}~\bibnamefont {Yamauchi}}, \bibinfo {author} {\bibfnamefont
  {H.}~\bibnamefont {Mimura}}, \bibinfo {author} {\bibfnamefont
  {H.}~\bibnamefont {Kitamura}}, \bibinfo {author} {\bibfnamefont
  {T.}~\bibnamefont {Katayama}}, \emph {et~al.},\ }\bibfield  {title} {\bibinfo
  {title} {Atomic inner-shell laser at 1.5-{\aa}ngstr{\"o}m wavelength pumped
  by an x-ray free-electron laser},\ }\href@noop {} {\bibfield  {journal}
  {\bibinfo  {journal} {Nature}\ }\textbf {\bibinfo {volume} {524}},\ \bibinfo
  {pages} {446} (\bibinfo {year} {2015})}\BibitemShut {NoStop}%
\bibitem [{\citenamefont {Comin}\ and\ \citenamefont
  {Damascelli}(2016)}]{comin2016resonant}%
  \BibitemOpen
  \bibfield  {author} {\bibinfo {author} {\bibfnamefont {R.}~\bibnamefont
  {Comin}}\ and\ \bibinfo {author} {\bibfnamefont {A.}~\bibnamefont
  {Damascelli}},\ }\bibfield  {title} {\bibinfo {title} {Resonant x-ray
  scattering studies of charge order in cuprates},\ }\href@noop {} {\bibfield
  {journal} {\bibinfo  {journal} {Annual Review of Condensed Matter Physics}\
  }\textbf {\bibinfo {volume} {7}},\ \bibinfo {pages} {369} (\bibinfo {year}
  {2016})}\BibitemShut {NoStop}%
\bibitem [{\citenamefont {Rouxel}\ \emph {et~al.}(2021)\citenamefont {Rouxel},
  \citenamefont {Keefer},\ and\ \citenamefont
  {Mukamel}}]{rouxel2021signatures}%
  \BibitemOpen
  \bibfield  {author} {\bibinfo {author} {\bibfnamefont {J.~R.}\ \bibnamefont
  {Rouxel}}, \bibinfo {author} {\bibfnamefont {D.}~\bibnamefont {Keefer}},\
  and\ \bibinfo {author} {\bibfnamefont {S.}~\bibnamefont {Mukamel}},\
  }\bibfield  {title} {\bibinfo {title} {Signatures of electronic and nuclear
  coherences in ultrafast molecular x-ray and electron diffraction},\
  }\href@noop {} {\bibfield  {journal} {\bibinfo  {journal} {Structural
  Dynamics}\ }\textbf {\bibinfo {volume} {8}},\ \bibinfo {pages} {014101}
  (\bibinfo {year} {2021})}\BibitemShut {NoStop}%
\bibitem [{\citenamefont {Dixit}\ \emph {et~al.}(2012)\citenamefont {Dixit},
  \citenamefont {Vendrell},\ and\ \citenamefont {Santra}}]{dixit2012imaging}%
  \BibitemOpen
  \bibfield  {author} {\bibinfo {author} {\bibfnamefont {G.}~\bibnamefont
  {Dixit}}, \bibinfo {author} {\bibfnamefont {O.}~\bibnamefont {Vendrell}},\
  and\ \bibinfo {author} {\bibfnamefont {R.}~\bibnamefont {Santra}},\
  }\bibfield  {title} {\bibinfo {title} {Imaging electronic quantum motion with
  light},\ }\href@noop {} {\bibfield  {journal} {\bibinfo  {journal}
  {Proceedings of the National Academy of Sciences}\ }\textbf {\bibinfo
  {volume} {109}},\ \bibinfo {pages} {11636} (\bibinfo {year}
  {2012})}\BibitemShut {NoStop}%
\bibitem [{\citenamefont {Feldhaus}\ \emph {et~al.}(2005)\citenamefont
  {Feldhaus}, \citenamefont {Arthur},\ and\ \citenamefont
  {Hastings}}]{feldhaus2005x}%
  \BibitemOpen
  \bibfield  {author} {\bibinfo {author} {\bibfnamefont {J.}~\bibnamefont
  {Feldhaus}}, \bibinfo {author} {\bibfnamefont {J.}~\bibnamefont {Arthur}},\
  and\ \bibinfo {author} {\bibfnamefont {J.}~\bibnamefont {Hastings}},\
  }\bibfield  {title} {\bibinfo {title} {X-ray free-electron lasers},\
  }\href@noop {} {\bibfield  {journal} {\bibinfo  {journal} {Journal of Physics
  B: Atomic, molecular and optical physics}\ }\textbf {\bibinfo {volume}
  {38}},\ \bibinfo {pages} {S799} (\bibinfo {year} {2005})}\BibitemShut
  {NoStop}%
\bibitem [{\citenamefont {Mukamel}\ \emph {et~al.}(2013)\citenamefont
  {Mukamel}, \citenamefont {Healion}, \citenamefont {Zhang},\ and\
  \citenamefont {Biggs}}]{mukamel2013multidimensional}%
  \BibitemOpen
  \bibfield  {author} {\bibinfo {author} {\bibfnamefont {S.}~\bibnamefont
  {Mukamel}}, \bibinfo {author} {\bibfnamefont {D.}~\bibnamefont {Healion}},
  \bibinfo {author} {\bibfnamefont {Y.}~\bibnamefont {Zhang}},\ and\ \bibinfo
  {author} {\bibfnamefont {J.~D.}\ \bibnamefont {Biggs}},\ }\bibfield  {title}
  {\bibinfo {title} {Multidimensional attosecond resonant x-ray spectroscopy of
  molecules: Lessons from the optical regime},\ }\href@noop {} {\bibfield
  {journal} {\bibinfo  {journal} {Annual Review of Physical Chemistry}\
  }\textbf {\bibinfo {volume} {64}},\ \bibinfo {pages} {101} (\bibinfo {year}
  {2013})}\BibitemShut {NoStop}%
\bibitem [{\citenamefont {Cornacchia}\ \emph {et~al.}(2004)\citenamefont
  {Cornacchia}, \citenamefont {Arthur}, \citenamefont {Bane}, \citenamefont
  {Bolton}, \citenamefont {Carr}, \citenamefont {Decker}, \citenamefont {Emma},
  \citenamefont {Galayda}, \citenamefont {Hastings}, \citenamefont {Hodgson}
  \emph {et~al.}}]{cornacchia2004future}%
  \BibitemOpen
  \bibfield  {author} {\bibinfo {author} {\bibfnamefont {M.}~\bibnamefont
  {Cornacchia}}, \bibinfo {author} {\bibfnamefont {J.}~\bibnamefont {Arthur}},
  \bibinfo {author} {\bibfnamefont {K.}~\bibnamefont {Bane}}, \bibinfo {author}
  {\bibfnamefont {P.}~\bibnamefont {Bolton}}, \bibinfo {author} {\bibfnamefont
  {R.}~\bibnamefont {Carr}}, \bibinfo {author} {\bibfnamefont {F.}~\bibnamefont
  {Decker}}, \bibinfo {author} {\bibfnamefont {P.}~\bibnamefont {Emma}},
  \bibinfo {author} {\bibfnamefont {J.}~\bibnamefont {Galayda}}, \bibinfo
  {author} {\bibfnamefont {J.}~\bibnamefont {Hastings}}, \bibinfo {author}
  {\bibfnamefont {K.}~\bibnamefont {Hodgson}}, \emph {et~al.},\ }\bibfield
  {title} {\bibinfo {title} {Future possibilities of the linac coherent light
  source},\ }\href@noop {} {\bibfield  {journal} {\bibinfo  {journal} {Journal
  of synchrotron radiation}\ }\textbf {\bibinfo {volume} {11}},\ \bibinfo
  {pages} {227} (\bibinfo {year} {2004})}\BibitemShut {NoStop}%
\bibitem [{\citenamefont {Bennett}\ \emph {et~al.}(2014)\citenamefont
  {Bennett}, \citenamefont {Biggs}, \citenamefont {Zhang}, \citenamefont
  {Dorfman},\ and\ \citenamefont {Mukamel}}]{bennett2014time}%
  \BibitemOpen
  \bibfield  {author} {\bibinfo {author} {\bibfnamefont {K.}~\bibnamefont
  {Bennett}}, \bibinfo {author} {\bibfnamefont {J.~D.}\ \bibnamefont {Biggs}},
  \bibinfo {author} {\bibfnamefont {Y.}~\bibnamefont {Zhang}}, \bibinfo
  {author} {\bibfnamefont {K.~E.}\ \bibnamefont {Dorfman}},\ and\ \bibinfo
  {author} {\bibfnamefont {S.}~\bibnamefont {Mukamel}},\ }\bibfield  {title}
  {\bibinfo {title} {Time-, frequency-, and wavevector-resolved x-ray
  diffraction from single molecules},\ }\href@noop {} {\bibfield  {journal}
  {\bibinfo  {journal} {The Journal of Chemical Physics}\ }\textbf {\bibinfo
  {volume} {140}},\ \bibinfo {pages} {204311} (\bibinfo {year}
  {2014})}\BibitemShut {NoStop}%
\bibitem [{\citenamefont {Als-Nielsen}\ and\ \citenamefont
  {McMorrow}(2011)}]{als2011elements}%
  \BibitemOpen
  \bibfield  {author} {\bibinfo {author} {\bibfnamefont {J.}~\bibnamefont
  {Als-Nielsen}}\ and\ \bibinfo {author} {\bibfnamefont {D.}~\bibnamefont
  {McMorrow}},\ }\href {https://books.google.de/books?id=fvrVDwAAQBAJ} {\emph
  {\bibinfo {title} {Elements of Modern X-ray Physics}}}\ (\bibinfo
  {publisher} {Wiley},\ \bibinfo {address} {Chichester},\ \bibinfo {year}
  {2011})\BibitemShut {NoStop}%
\bibitem [{\citenamefont {Helliwell}(1984)}]{helliwell1984synchrotron}%
  \BibitemOpen
  \bibfield  {author} {\bibinfo {author} {\bibfnamefont {J.}~\bibnamefont
  {Helliwell}},\ }\bibfield  {title} {\bibinfo {title} {Synchrotron x-radiation
  protein crystallography: instrumentation, methods and applications},\
  }\href@noop {} {\bibfield  {journal} {\bibinfo  {journal} {Reports on
  Progress in Physics}\ }\textbf {\bibinfo {volume} {47}},\ \bibinfo {pages}
  {1403} (\bibinfo {year} {1984})}\BibitemShut {NoStop}%
\bibitem [{\citenamefont {Freund}(1989)}]{freund1989x}%
  \BibitemOpen
  \bibfield  {author} {\bibinfo {author} {\bibfnamefont {A.~K.}\ \bibnamefont
  {Freund}},\ }\bibfield  {title} {\bibinfo {title} {X-ray optics for
  synchrotron radiation},\ }in\ \href@noop {} {\emph {\bibinfo {booktitle}
  {Synchrotron Radiation in Structural Biology}}}\ (\bibinfo  {publisher}
  {Springer US},\ \bibinfo {address} {Boston, MA},\ \bibinfo {year} {1989})\
  pp.\ \bibinfo {pages} {255--292}\BibitemShut {NoStop}%
\bibitem [{\citenamefont {Cao}\ and\ \citenamefont
  {Wilson}(1998)}]{cao1998ultrafast}%
  \BibitemOpen
  \bibfield  {author} {\bibinfo {author} {\bibfnamefont {J.}~\bibnamefont
  {Cao}}\ and\ \bibinfo {author} {\bibfnamefont {K.~R.}\ \bibnamefont
  {Wilson}},\ }\bibfield  {title} {\bibinfo {title} {Ultrafast x-ray
  diffraction theory},\ }\href@noop {} {\bibfield  {journal} {\bibinfo
  {journal} {The Journal of Physical Chemistry A}\ }\textbf {\bibinfo {volume}
  {102}},\ \bibinfo {pages} {9523} (\bibinfo {year} {1998})}\BibitemShut
  {NoStop}%
\bibitem [{\citenamefont {Ben-Nun}\ \emph {et~al.}(1997)\citenamefont
  {Ben-Nun}, \citenamefont {Cao},\ and\ \citenamefont
  {Wilson}}]{ben1997ultrafast}%
  \BibitemOpen
  \bibfield  {author} {\bibinfo {author} {\bibfnamefont {M.}~\bibnamefont
  {Ben-Nun}}, \bibinfo {author} {\bibfnamefont {J.}~\bibnamefont {Cao}},\ and\
  \bibinfo {author} {\bibfnamefont {K.~R.}\ \bibnamefont {Wilson}},\ }\bibfield
   {title} {\bibinfo {title} {Ultrafast x-ray and electron diffraction:
  theoretical considerations},\ }\href@noop {} {\bibfield  {journal} {\bibinfo
  {journal} {The Journal of Physical Chemistry A}\ }\textbf {\bibinfo {volume}
  {101}},\ \bibinfo {pages} {8743} (\bibinfo {year} {1997})}\BibitemShut
  {NoStop}%
\bibitem [{\citenamefont {Dashti}\ \emph {et~al.}(2020)\citenamefont {Dashti},
  \citenamefont {Mashayekhi}, \citenamefont {Shekhar}, \citenamefont {Hail},
  \citenamefont {Salah}, \citenamefont {Schwander}, \citenamefont {des
  Georges}, \citenamefont {Singharoy}, \citenamefont {Frank},\ and\
  \citenamefont {Ourmazd}}]{dashti2020retrieving}%
  \BibitemOpen
  \bibfield  {author} {\bibinfo {author} {\bibfnamefont {A.}~\bibnamefont
  {Dashti}}, \bibinfo {author} {\bibfnamefont {G.}~\bibnamefont {Mashayekhi}},
  \bibinfo {author} {\bibfnamefont {M.}~\bibnamefont {Shekhar}}, \bibinfo
  {author} {\bibfnamefont {D.~B.}\ \bibnamefont {Hail}}, \bibinfo {author}
  {\bibfnamefont {S.}~\bibnamefont {Salah}}, \bibinfo {author} {\bibfnamefont
  {P.}~\bibnamefont {Schwander}}, \bibinfo {author} {\bibfnamefont
  {A.}~\bibnamefont {des Georges}}, \bibinfo {author} {\bibfnamefont
  {A.}~\bibnamefont {Singharoy}}, \bibinfo {author} {\bibfnamefont
  {J.}~\bibnamefont {Frank}},\ and\ \bibinfo {author} {\bibfnamefont
  {A.}~\bibnamefont {Ourmazd}},\ }\bibfield  {title} {\bibinfo {title}
  {Retrieving functional pathways of biomolecules from single-particle
  snapshots},\ }\href@noop {} {\bibfield  {journal} {\bibinfo  {journal}
  {Nature Communications}\ }\textbf {\bibinfo {volume} {11}},\ \bibinfo {pages}
  {1} (\bibinfo {year} {2020})}\BibitemShut {NoStop}%
\bibitem [{\citenamefont {Dixit}\ and\ \citenamefont
  {Santra}(2017)}]{dixit2017time}%
  \BibitemOpen
  \bibfield  {author} {\bibinfo {author} {\bibfnamefont {G.}~\bibnamefont
  {Dixit}}\ and\ \bibinfo {author} {\bibfnamefont {R.}~\bibnamefont {Santra}},\
  }\bibfield  {title} {\bibinfo {title} {Time-resolved ultrafast x-ray
  scattering from an incoherent electronic mixture},\ }\href@noop {} {\bibfield
   {journal} {\bibinfo  {journal} {Physical Review A}\ }\textbf {\bibinfo
  {volume} {96}},\ \bibinfo {pages} {053413} (\bibinfo {year}
  {2017})}\BibitemShut {NoStop}%
\bibitem [{\citenamefont {Coridan}\ \emph {et~al.}(2012)\citenamefont
  {Coridan}, \citenamefont {Schmidt}, \citenamefont {Lai}, \citenamefont
  {Abbamonte},\ and\ \citenamefont {Wong}}]{coridan2012dynamics}%
  \BibitemOpen
  \bibfield  {author} {\bibinfo {author} {\bibfnamefont {R.~H.}\ \bibnamefont
  {Coridan}}, \bibinfo {author} {\bibfnamefont {N.~W.}\ \bibnamefont
  {Schmidt}}, \bibinfo {author} {\bibfnamefont {G.~H.}\ \bibnamefont {Lai}},
  \bibinfo {author} {\bibfnamefont {P.}~\bibnamefont {Abbamonte}},\ and\
  \bibinfo {author} {\bibfnamefont {G.~C.}\ \bibnamefont {Wong}},\ }\bibfield
  {title} {\bibinfo {title} {Dynamics of confined water reconstructed from
  inelastic x-ray scattering measurements of bulk response functions},\
  }\href@noop {} {\bibfield  {journal} {\bibinfo  {journal} {Physical Review
  E}\ }\textbf {\bibinfo {volume} {85}},\ \bibinfo {pages} {031501} (\bibinfo
  {year} {2012})}\BibitemShut {NoStop}%
\bibitem [{\citenamefont {Asban}\ \emph {et~al.}(2019)\citenamefont {Asban},
  \citenamefont {Dorfman},\ and\ \citenamefont {Mukamel}}]{asban2019quantum}%
  \BibitemOpen
  \bibfield  {author} {\bibinfo {author} {\bibfnamefont {S.}~\bibnamefont
  {Asban}}, \bibinfo {author} {\bibfnamefont {K.~E.}\ \bibnamefont {Dorfman}},\
  and\ \bibinfo {author} {\bibfnamefont {S.}~\bibnamefont {Mukamel}},\
  }\bibfield  {title} {\bibinfo {title} {Quantum phase-sensitive diffraction
  and imaging using entangled photons},\ }\href@noop {} {\bibfield  {journal}
  {\bibinfo  {journal} {Proceedings of the National Academy of Sciences}\
  }\textbf {\bibinfo {volume} {116}},\ \bibinfo {pages} {11673} (\bibinfo
  {year} {2019})}\BibitemShut {NoStop}%
\bibitem [{\citenamefont {Craig}\ and\ \citenamefont
  {Thirunamachandran}(1998)}]{craig1998molecular}%
  \BibitemOpen
  \bibfield  {author} {\bibinfo {author} {\bibfnamefont {D.~P.}\ \bibnamefont
  {Craig}}\ and\ \bibinfo {author} {\bibfnamefont {T.}~\bibnamefont
  {Thirunamachandran}},\ }\href@noop {} {\emph {\bibinfo {title} {Molecular
  quantum electrodynamics: an introduction to radiation-molecule
  interactions}}}\ (\bibinfo  {publisher} {Dover Publications},\ \bibinfo
  {address} {New York},\ \bibinfo {year} {1998})\BibitemShut {NoStop}%
\bibitem [{\citenamefont {Fetter}\ and\ \citenamefont
  {Walecka}()}]{fetter1971quantum}%
  \BibitemOpen
  \bibfield  {author} {\bibinfo {author} {\bibfnamefont {A.}~\bibnamefont
  {Fetter}}\ and\ \bibinfo {author} {\bibfnamefont {J.}~\bibnamefont
  {Walecka}},\ }\href@noop {} {\emph {\bibinfo {title} {1971Quantum theory of
  many-particle systems}}}\ (\bibinfo  {publisher} {New York:
  McGraw-Hill})\BibitemShut {NoStop}%
\bibitem [{\citenamefont {Santra}(2008)}]{santra2008concepts}%
  \BibitemOpen
  \bibfield  {author} {\bibinfo {author} {\bibfnamefont {R.}~\bibnamefont
  {Santra}},\ }\bibfield  {title} {\bibinfo {title} {Concepts in x-ray
  physics},\ }\href@noop {} {\bibfield  {journal} {\bibinfo  {journal} {Journal
  of Physics B: Atomic, Molecular and Optical Physics}\ }\textbf {\bibinfo
  {volume} {42}},\ \bibinfo {pages} {023001} (\bibinfo {year}
  {2008})}\BibitemShut {NoStop}%
\bibitem [{\citenamefont {Lorenz}\ \emph {et~al.}(2012)\citenamefont {Lorenz},
  \citenamefont {Kabachnik}, \citenamefont {Weckert},\ and\ \citenamefont
  {Vartanyants}}]{lorenz2012impact}%
  \BibitemOpen
  \bibfield  {author} {\bibinfo {author} {\bibfnamefont {U.}~\bibnamefont
  {Lorenz}}, \bibinfo {author} {\bibfnamefont {N.}~\bibnamefont {Kabachnik}},
  \bibinfo {author} {\bibfnamefont {E.}~\bibnamefont {Weckert}},\ and\ \bibinfo
  {author} {\bibfnamefont {I.}~\bibnamefont {Vartanyants}},\ }\bibfield
  {title} {\bibinfo {title} {Impact of ultrafast electronic damage in
  single-particle x-ray imaging experiments},\ }\href@noop {} {\bibfield
  {journal} {\bibinfo  {journal} {Physical Review E}\ }\textbf {\bibinfo
  {volume} {86}},\ \bibinfo {pages} {051911} (\bibinfo {year}
  {2012})}\BibitemShut {NoStop}%
\bibitem [{\citenamefont {Keldysh}\ \emph {et~al.}(1965)\citenamefont {Keldysh}
  \emph {et~al.}}]{keldysh1965diagram}%
  \BibitemOpen
  \bibfield  {author} {\bibinfo {author} {\bibfnamefont {L.~V.}\ \bibnamefont
  {Keldysh}} \emph {et~al.},\ }\bibfield  {title} {\bibinfo {title} {Diagram
  technique for nonequilibrium processes},\ }\href@noop {} {\bibfield
  {journal} {\bibinfo  {journal} {Sov. Phys. JETP}\ }\textbf {\bibinfo {volume}
  {20}},\ \bibinfo {pages} {1018} (\bibinfo {year} {1965})}\BibitemShut
  {NoStop}%
\bibitem [{\citenamefont {Fleischhauer}(1998)}]{fleischhauer1998quantum}%
  \BibitemOpen
  \bibfield  {author} {\bibinfo {author} {\bibfnamefont {M.}~\bibnamefont
  {Fleischhauer}},\ }\bibfield  {title} {\bibinfo {title} {Quantum-theory of
  photodetection without the rotating wave approximation},\ }\href@noop {}
  {\bibfield  {journal} {\bibinfo  {journal} {Journal of Physics A:
  Mathematical and General}\ }\textbf {\bibinfo {volume} {31}},\ \bibinfo
  {pages} {453} (\bibinfo {year} {1998})}\BibitemShut {NoStop}%
\bibitem [{\citenamefont {Mukamel}(2003)}]{mukamel2003superoperator}%
  \BibitemOpen
  \bibfield  {author} {\bibinfo {author} {\bibfnamefont {S.}~\bibnamefont
  {Mukamel}},\ }\bibfield  {title} {\bibinfo {title} {Superoperator
  representation of nonlinear response: Unifying quantum field and mode
  coupling theories},\ }\href@noop {} {\bibfield  {journal} {\bibinfo
  {journal} {Physical Review E}\ }\textbf {\bibinfo {volume} {68}},\ \bibinfo
  {pages} {021111} (\bibinfo {year} {2003})}\BibitemShut {NoStop}%
\bibitem [{\citenamefont {Dorfman}\ and\ \citenamefont
  {Mukamel}(2016)}]{dorfman2016time}%
  \BibitemOpen
  \bibfield  {author} {\bibinfo {author} {\bibfnamefont {K.~E.}\ \bibnamefont
  {Dorfman}}\ and\ \bibinfo {author} {\bibfnamefont {S.}~\bibnamefont
  {Mukamel}},\ }\bibfield  {title} {\bibinfo {title} {Time-and-frequency-gated
  photon coincidence counting; a novel multidimensional spectroscopy tool},\
  }\href@noop {} {\bibfield  {journal} {\bibinfo  {journal} {Physica Scripta}\
  }\textbf {\bibinfo {volume} {91}},\ \bibinfo {pages} {083004} (\bibinfo
  {year} {2016})}\BibitemShut {NoStop}%
\bibitem [{\citenamefont {Dorfman}\ and\ \citenamefont
  {Mukamel}(2012)}]{dorfman2012nonlinear}%
  \BibitemOpen
  \bibfield  {author} {\bibinfo {author} {\bibfnamefont {K.~E.}\ \bibnamefont
  {Dorfman}}\ and\ \bibinfo {author} {\bibfnamefont {S.}~\bibnamefont
  {Mukamel}},\ }\bibfield  {title} {\bibinfo {title} {Nonlinear spectroscopy
  with time-and frequency-gated photon counting: A superoperator diagrammatic
  approach},\ }\href@noop {} {\bibfield  {journal} {\bibinfo  {journal}
  {Physical Review A}\ }\textbf {\bibinfo {volume} {86}},\ \bibinfo {pages}
  {013810} (\bibinfo {year} {2012})}\BibitemShut {NoStop}%
\bibitem [{\citenamefont {Vartanyants}\ and\ \citenamefont
  {Singer}(2010)}]{vartanyants2010coherence}%
  \BibitemOpen
  \bibfield  {author} {\bibinfo {author} {\bibfnamefont {I.~A.}\ \bibnamefont
  {Vartanyants}}\ and\ \bibinfo {author} {\bibfnamefont {A.}~\bibnamefont
  {Singer}},\ }\bibfield  {title} {\bibinfo {title} {Coherence properties of
  hard x-ray synchrotron sources and x-ray free-electron lasers},\ }\href@noop
  {} {\bibfield  {journal} {\bibinfo  {journal} {New Journal of Physics}\
  }\textbf {\bibinfo {volume} {12}},\ \bibinfo {pages} {035004} (\bibinfo
  {year} {2010})}\BibitemShut {NoStop}%
\bibitem [{\citenamefont {Pietsch}\ \emph {et~al.}(2005)\citenamefont
  {Pietsch}, \citenamefont {Panzner}, \citenamefont {Leitenberger},\ and\
  \citenamefont {Vartanyants}}]{pietsch2005coherence}%
  \BibitemOpen
  \bibfield  {author} {\bibinfo {author} {\bibfnamefont {U.}~\bibnamefont
  {Pietsch}}, \bibinfo {author} {\bibfnamefont {T.}~\bibnamefont {Panzner}},
  \bibinfo {author} {\bibfnamefont {W.}~\bibnamefont {Leitenberger}},\ and\
  \bibinfo {author} {\bibfnamefont {I.}~\bibnamefont {Vartanyants}},\
  }\bibfield  {title} {\bibinfo {title} {Coherence experiments using white
  synchrotron radiation},\ }\href@noop {} {\bibfield  {journal} {\bibinfo
  {journal} {Physica B: Condensed Matter}\ }\textbf {\bibinfo {volume} {357}},\
  \bibinfo {pages} {45} (\bibinfo {year} {2005})}\BibitemShut {NoStop}%
\bibitem [{\citenamefont {Skopintsev}\ \emph {et~al.}(2014)\citenamefont
  {Skopintsev}, \citenamefont {Singer}, \citenamefont {Bach}, \citenamefont
  {M{\"u}ller}, \citenamefont {Beyersdorff}, \citenamefont {Schleitzer},
  \citenamefont {Gorobtsov}, \citenamefont {Shabalin}, \citenamefont {Kurta},
  \citenamefont {Dzhigaev} \emph {et~al.}}]{skopintsev2014characterization}%
  \BibitemOpen
  \bibfield  {author} {\bibinfo {author} {\bibfnamefont {P.}~\bibnamefont
  {Skopintsev}}, \bibinfo {author} {\bibfnamefont {A.}~\bibnamefont {Singer}},
  \bibinfo {author} {\bibfnamefont {J.}~\bibnamefont {Bach}}, \bibinfo {author}
  {\bibfnamefont {L.}~\bibnamefont {M{\"u}ller}}, \bibinfo {author}
  {\bibfnamefont {B.}~\bibnamefont {Beyersdorff}}, \bibinfo {author}
  {\bibfnamefont {S.}~\bibnamefont {Schleitzer}}, \bibinfo {author}
  {\bibfnamefont {O.}~\bibnamefont {Gorobtsov}}, \bibinfo {author}
  {\bibfnamefont {A.}~\bibnamefont {Shabalin}}, \bibinfo {author}
  {\bibfnamefont {R.}~\bibnamefont {Kurta}}, \bibinfo {author} {\bibfnamefont
  {D.}~\bibnamefont {Dzhigaev}}, \emph {et~al.},\ }\bibfield  {title} {\bibinfo
  {title} {Characterization of spatial coherence of synchrotron radiation with
  non-redundant arrays of apertures},\ }\href@noop {} {\bibfield  {journal}
  {\bibinfo  {journal} {Journal of synchrotron radiation}\ }\textbf {\bibinfo
  {volume} {21}},\ \bibinfo {pages} {722} (\bibinfo {year} {2014})}\BibitemShut
  {NoStop}%
\bibitem [{\citenamefont {Khubbutdinov}\ \emph {et~al.}(2019)\citenamefont
  {Khubbutdinov}, \citenamefont {Menushenkov},\ and\ \citenamefont
  {Vartanyants}}]{khubbutdinov2019coherence}%
  \BibitemOpen
  \bibfield  {author} {\bibinfo {author} {\bibfnamefont {R.}~\bibnamefont
  {Khubbutdinov}}, \bibinfo {author} {\bibfnamefont {A.}~\bibnamefont
  {Menushenkov}},\ and\ \bibinfo {author} {\bibfnamefont {I.}~\bibnamefont
  {Vartanyants}},\ }\bibfield  {title} {\bibinfo {title} {Coherence properties
  of the high-energy fourth-generation x-ray synchrotron sources},\ }\href@noop
  {} {\bibfield  {journal} {\bibinfo  {journal} {Journal of synchrotron
  radiation}\ }\textbf {\bibinfo {volume} {26}},\ \bibinfo {pages} {1851}
  (\bibinfo {year} {2019})}\BibitemShut {NoStop}%
\bibitem [{\citenamefont {Van~Hove}(1954)}]{van1954correlations}%
  \BibitemOpen
  \bibfield  {author} {\bibinfo {author} {\bibfnamefont {L.}~\bibnamefont
  {Van~Hove}},\ }\bibfield  {title} {\bibinfo {title} {Correlations in space
  and time and born approximation scattering in systems of interacting
  particles},\ }\href@noop {} {\bibfield  {journal} {\bibinfo  {journal}
  {Physical Review}\ }\textbf {\bibinfo {volume} {95}},\ \bibinfo {pages} {249}
  (\bibinfo {year} {1954})}\BibitemShut {NoStop}%
\bibitem [{\citenamefont {Abbamonte}\ \emph {et~al.}(2012)\citenamefont
  {Abbamonte}, \citenamefont {Demler}, \citenamefont {Davis},\ and\
  \citenamefont {Campuzano}}]{abbamonte2012resonant}%
  \BibitemOpen
  \bibfield  {author} {\bibinfo {author} {\bibfnamefont {P.}~\bibnamefont
  {Abbamonte}}, \bibinfo {author} {\bibfnamefont {E.}~\bibnamefont {Demler}},
  \bibinfo {author} {\bibfnamefont {J.~S.}\ \bibnamefont {Davis}},\ and\
  \bibinfo {author} {\bibfnamefont {J.-C.}\ \bibnamefont {Campuzano}},\
  }\bibfield  {title} {\bibinfo {title} {Resonant soft x-ray scattering, stripe
  order, and the electron spectral function in cuprates},\ }\href@noop {}
  {\bibfield  {journal} {\bibinfo  {journal} {Physica C: Superconductivity}\
  }\textbf {\bibinfo {volume} {481}},\ \bibinfo {pages} {15} (\bibinfo {year}
  {2012})}\BibitemShut {NoStop}%
\bibitem [{\citenamefont {Gan}\ \emph {et~al.}(2013)\citenamefont {Gan},
  \citenamefont {Kogar},\ and\ \citenamefont
  {Abbamonte}}]{gan2013crystallographic}%
  \BibitemOpen
  \bibfield  {author} {\bibinfo {author} {\bibfnamefont {Y.}~\bibnamefont
  {Gan}}, \bibinfo {author} {\bibfnamefont {A.}~\bibnamefont {Kogar}},\ and\
  \bibinfo {author} {\bibfnamefont {P.}~\bibnamefont {Abbamonte}},\ }\bibfield
  {title} {\bibinfo {title} {Crystallographic refinement of collective
  excitations using standing wave inelastic x-ray scattering},\ }\href@noop {}
  {\bibfield  {journal} {\bibinfo  {journal} {Chemical Physics}\ }\textbf
  {\bibinfo {volume} {414}},\ \bibinfo {pages} {160} (\bibinfo {year}
  {2013})}\BibitemShut {NoStop}%
\bibitem [{\citenamefont {Cavalleri}\ \emph {et~al.}(2006)\citenamefont
  {Cavalleri}, \citenamefont {Wall}, \citenamefont {Simpson}, \citenamefont
  {Statz}, \citenamefont {Ward}, \citenamefont {Nelson}, \citenamefont {Rini},\
  and\ \citenamefont {Schoenlein}}]{cavalleri2006tracking}%
  \BibitemOpen
  \bibfield  {author} {\bibinfo {author} {\bibfnamefont {A.}~\bibnamefont
  {Cavalleri}}, \bibinfo {author} {\bibfnamefont {S.}~\bibnamefont {Wall}},
  \bibinfo {author} {\bibfnamefont {C.}~\bibnamefont {Simpson}}, \bibinfo
  {author} {\bibfnamefont {E.}~\bibnamefont {Statz}}, \bibinfo {author}
  {\bibfnamefont {D.}~\bibnamefont {Ward}}, \bibinfo {author} {\bibfnamefont
  {K.}~\bibnamefont {Nelson}}, \bibinfo {author} {\bibfnamefont
  {M.}~\bibnamefont {Rini}},\ and\ \bibinfo {author} {\bibfnamefont
  {R.}~\bibnamefont {Schoenlein}},\ }\bibfield  {title} {\bibinfo {title}
  {Tracking the motion of charges in a terahertz light field by femtosecond
  x-ray diffraction},\ }\href@noop {} {\bibfield  {journal} {\bibinfo
  {journal} {Nature}\ }\textbf {\bibinfo {volume} {442}},\ \bibinfo {pages}
  {664} (\bibinfo {year} {2006})}\BibitemShut {NoStop}%
\bibitem [{\citenamefont {Huang}\ and\ \citenamefont
  {Lucas}(2021)}]{huang2021electron}%
  \BibitemOpen
  \bibfield  {author} {\bibinfo {author} {\bibfnamefont {X.}~\bibnamefont
  {Huang}}\ and\ \bibinfo {author} {\bibfnamefont {A.}~\bibnamefont {Lucas}},\
  }\bibfield  {title} {\bibinfo {title} {Electron-phonon hydrodynamics},\
  }\href@noop {} {\bibfield  {journal} {\bibinfo  {journal} {Physical Review
  B}\ }\textbf {\bibinfo {volume} {103}},\ \bibinfo {pages} {155128} (\bibinfo
  {year} {2021})}\BibitemShut {NoStop}%
\bibitem [{\citenamefont {Tabis}\ \emph {et~al.}(2017)\citenamefont {Tabis},
  \citenamefont {Yu}, \citenamefont {Bialo}, \citenamefont {Bluschke},
  \citenamefont {Kolodziej}, \citenamefont {Kozlowski}, \citenamefont
  {Blackburn}, \citenamefont {Sen}, \citenamefont {Forgan}, \citenamefont
  {Zimmermann} \emph {et~al.}}]{tabis2017synchrotron}%
  \BibitemOpen
  \bibfield  {author} {\bibinfo {author} {\bibfnamefont {W.}~\bibnamefont
  {Tabis}}, \bibinfo {author} {\bibfnamefont {B.}~\bibnamefont {Yu}}, \bibinfo
  {author} {\bibfnamefont {I.}~\bibnamefont {Bialo}}, \bibinfo {author}
  {\bibfnamefont {M.}~\bibnamefont {Bluschke}}, \bibinfo {author}
  {\bibfnamefont {T.}~\bibnamefont {Kolodziej}}, \bibinfo {author}
  {\bibfnamefont {A.}~\bibnamefont {Kozlowski}}, \bibinfo {author}
  {\bibfnamefont {E.}~\bibnamefont {Blackburn}}, \bibinfo {author}
  {\bibfnamefont {K.}~\bibnamefont {Sen}}, \bibinfo {author} {\bibfnamefont
  {E.~M.}\ \bibnamefont {Forgan}}, \bibinfo {author} {\bibfnamefont {M.~v.}\
  \bibnamefont {Zimmermann}}, \emph {et~al.},\ }\bibfield  {title} {\bibinfo
  {title} {Synchrotron x-ray scattering study of charge-density-wave order in
  hgba 2 cuo 4+ $\delta$},\ }\href@noop {} {\bibfield  {journal} {\bibinfo
  {journal} {Physical Review B}\ }\textbf {\bibinfo {volume} {96}},\ \bibinfo
  {pages} {134510} (\bibinfo {year} {2017})}\BibitemShut {NoStop}%
\bibitem [{\citenamefont {Hartnoll}\ and\ \citenamefont
  {Mackenzie}(2022)}]{hartnoll2022colloquium}%
  \BibitemOpen
  \bibfield  {author} {\bibinfo {author} {\bibfnamefont {S.~A.}\ \bibnamefont
  {Hartnoll}}\ and\ \bibinfo {author} {\bibfnamefont {A.~P.}\ \bibnamefont
  {Mackenzie}},\ }\bibfield  {title} {\bibinfo {title} {Colloquium: Planckian
  dissipation in metals},\ }\href@noop {} {\bibfield  {journal} {\bibinfo
  {journal} {Reviews of Modern Physics}\ }\textbf {\bibinfo {volume} {94}},\
  \bibinfo {pages} {041002} (\bibinfo {year} {2022})}\BibitemShut {NoStop}%
\bibitem [{\citenamefont {Mousatov}\ \emph {et~al.}(2020)\citenamefont
  {Mousatov}, \citenamefont {Berg},\ and\ \citenamefont
  {Hartnoll}}]{mousatov2020theory}%
  \BibitemOpen
  \bibfield  {author} {\bibinfo {author} {\bibfnamefont {C.~H.}\ \bibnamefont
  {Mousatov}}, \bibinfo {author} {\bibfnamefont {E.}~\bibnamefont {Berg}},\
  and\ \bibinfo {author} {\bibfnamefont {S.~A.}\ \bibnamefont {Hartnoll}},\
  }\bibfield  {title} {\bibinfo {title} {Theory of the strange metal
  sr3ru2o7},\ }\href@noop {} {\bibfield  {journal} {\bibinfo  {journal}
  {Proceedings of the National Academy of Sciences}\ }\textbf {\bibinfo
  {volume} {117}},\ \bibinfo {pages} {2852} (\bibinfo {year}
  {2020})}\BibitemShut {NoStop}%
\bibitem [{\citenamefont {Thornton}\ \emph {et~al.}(2022)\citenamefont
  {Thornton}, \citenamefont {Liarte}, \citenamefont {Abbamonte}, \citenamefont
  {Sethna},\ and\ \citenamefont {Chowdhury}}]{thornton2022jamming}%
  \BibitemOpen
  \bibfield  {author} {\bibinfo {author} {\bibfnamefont {S.~J.}\ \bibnamefont
  {Thornton}}, \bibinfo {author} {\bibfnamefont {D.~B.}\ \bibnamefont
  {Liarte}}, \bibinfo {author} {\bibfnamefont {P.}~\bibnamefont {Abbamonte}},
  \bibinfo {author} {\bibfnamefont {J.~P.}\ \bibnamefont {Sethna}},\ and\
  \bibinfo {author} {\bibfnamefont {D.}~\bibnamefont {Chowdhury}},\ }\bibfield
  {title} {\bibinfo {title} {Jamming and unusual charge density fluctuations of
  strange metals},\ }\href@noop {} {\bibfield  {journal} {\bibinfo  {journal}
  {arXiv:2210.16325}\ } (\bibinfo {year} {2022})}\BibitemShut {NoStop}%
\bibitem [{\citenamefont {Debnath}\ and\ \citenamefont
  {Rubio}(2020)}]{debnath2020entangled}%
  \BibitemOpen
  \bibfield  {author} {\bibinfo {author} {\bibfnamefont {A.}~\bibnamefont
  {Debnath}}\ and\ \bibinfo {author} {\bibfnamefont {A.}~\bibnamefont
  {Rubio}},\ }\bibfield  {title} {\bibinfo {title} {Entangled photon assisted
  multidimensional nonlinear optics of exciton--polaritons},\ }\href@noop {}
  {\bibfield  {journal} {\bibinfo  {journal} {Journal of Applied Physics}\
  }\textbf {\bibinfo {volume} {128}},\ \bibinfo {pages} {113102} (\bibinfo
  {year} {2020})}\BibitemShut {NoStop}%
\bibitem [{\citenamefont {Debnath}\ and\ \citenamefont
  {Rubio}(2022)}]{10.3389/fphy.2022.879113}%
  \BibitemOpen
  \bibfield  {author} {\bibinfo {author} {\bibfnamefont {A.}~\bibnamefont
  {Debnath}}\ and\ \bibinfo {author} {\bibfnamefont {A.}~\bibnamefont
  {Rubio}},\ }\bibfield  {title} {\bibinfo {title} {Entangled biphoton enhanced
  double quantum coherence signal as a probe for cavity polariton correlations
  in presence of phonon induced dephasing},\ }\href@noop {} {\bibfield
  {journal} {\bibinfo  {journal} {Frontiers in Physics}\ }\textbf {\bibinfo
  {volume} {10}} (\bibinfo {year} {2022})}\BibitemShut {NoStop}%
\bibitem [{\citenamefont {Bonitz}\ \emph {et~al.}(2020)\citenamefont {Bonitz},
  \citenamefont {Dornheim}, \citenamefont {Moldabekov}, \citenamefont {Zhang},
  \citenamefont {Hamann}, \citenamefont {K{\"a}hlert}, \citenamefont {Filinov},
  \citenamefont {Ramakrishna},\ and\ \citenamefont {Vorberger}}]{bonitz2020ab}%
  \BibitemOpen
  \bibfield  {author} {\bibinfo {author} {\bibfnamefont {M.}~\bibnamefont
  {Bonitz}}, \bibinfo {author} {\bibfnamefont {T.}~\bibnamefont {Dornheim}},
  \bibinfo {author} {\bibfnamefont {Z.~A.}\ \bibnamefont {Moldabekov}},
  \bibinfo {author} {\bibfnamefont {S.}~\bibnamefont {Zhang}}, \bibinfo
  {author} {\bibfnamefont {P.}~\bibnamefont {Hamann}}, \bibinfo {author}
  {\bibfnamefont {H.}~\bibnamefont {K{\"a}hlert}}, \bibinfo {author}
  {\bibfnamefont {A.}~\bibnamefont {Filinov}}, \bibinfo {author} {\bibfnamefont
  {K.}~\bibnamefont {Ramakrishna}},\ and\ \bibinfo {author} {\bibfnamefont
  {J.}~\bibnamefont {Vorberger}},\ }\bibfield  {title} {\bibinfo {title} {Ab
  initio simulation of warm dense matter},\ }\href@noop {} {\bibfield
  {journal} {\bibinfo  {journal} {Physics of Plasmas}\ }\textbf {\bibinfo
  {volume} {27}},\ \bibinfo {pages} {042710} (\bibinfo {year}
  {2020})}\BibitemShut {NoStop}%
\bibitem [{\citenamefont {Dornheim}\ \emph {et~al.}(2020)\citenamefont
  {Dornheim}, \citenamefont {Vorberger},\ and\ \citenamefont
  {Bonitz}}]{dornheim2020nonlinear}%
  \BibitemOpen
  \bibfield  {author} {\bibinfo {author} {\bibfnamefont {T.}~\bibnamefont
  {Dornheim}}, \bibinfo {author} {\bibfnamefont {J.}~\bibnamefont
  {Vorberger}},\ and\ \bibinfo {author} {\bibfnamefont {M.}~\bibnamefont
  {Bonitz}},\ }\bibfield  {title} {\bibinfo {title} {Nonlinear electronic
  density response in warm dense matter},\ }\href@noop {} {\bibfield  {journal}
  {\bibinfo  {journal} {Physical Review Letters}\ }\textbf {\bibinfo {volume}
  {125}},\ \bibinfo {pages} {085001} (\bibinfo {year} {2020})}\BibitemShut
  {NoStop}%
\bibitem [{\citenamefont {Dornheim}\ \emph {et~al.}(2021)\citenamefont
  {Dornheim}, \citenamefont {Moldabekov},\ and\ \citenamefont
  {Vorberger}}]{dornheim2021nonlinear}%
  \BibitemOpen
  \bibfield  {author} {\bibinfo {author} {\bibfnamefont {T.}~\bibnamefont
  {Dornheim}}, \bibinfo {author} {\bibfnamefont {Z.~A.}\ \bibnamefont
  {Moldabekov}},\ and\ \bibinfo {author} {\bibfnamefont {J.}~\bibnamefont
  {Vorberger}},\ }\bibfield  {title} {\bibinfo {title} {Nonlinear density
  response from imaginary-time correlation functions: Ab initio path integral
  monte carlo simulations of the warm dense electron gas},\ }\href@noop {}
  {\bibfield  {journal} {\bibinfo  {journal} {The Journal of Chemical Physics}\
  }\textbf {\bibinfo {volume} {155}},\ \bibinfo {pages} {054110} (\bibinfo
  {year} {2021})}\BibitemShut {NoStop}%
\bibitem [{\citenamefont {Dornheim}\ \emph {et~al.}(2022)\citenamefont
  {Dornheim}, \citenamefont {Vorberger}, \citenamefont {Moldabekov},\ and\
  \citenamefont {B{\"o}hme}}]{dornheim2022analyzing}%
  \BibitemOpen
  \bibfield  {author} {\bibinfo {author} {\bibfnamefont {T.}~\bibnamefont
  {Dornheim}}, \bibinfo {author} {\bibfnamefont {J.}~\bibnamefont {Vorberger}},
  \bibinfo {author} {\bibfnamefont {Z.}~\bibnamefont {Moldabekov}},\ and\
  \bibinfo {author} {\bibfnamefont {M.}~\bibnamefont {B{\"o}hme}},\ }\bibfield
  {title} {\bibinfo {title} {Analyzing x-ray thomson scattering experiments of
  warm dense matter in the imaginary-time domain: theoretical models and
  simulations},\ }\href@noop {} {\bibfield  {journal} {\bibinfo  {journal}
  {arXiv preprint arXiv:2211.00579}\ } (\bibinfo {year} {2022})}\BibitemShut
  {NoStop}%
\bibitem [{\citenamefont {Son}\ \emph {et~al.}(2011)\citenamefont {Son},
  \citenamefont {Young},\ and\ \citenamefont {Santra}}]{son2011impact}%
  \BibitemOpen
  \bibfield  {author} {\bibinfo {author} {\bibfnamefont {S.-K.}\ \bibnamefont
  {Son}}, \bibinfo {author} {\bibfnamefont {L.}~\bibnamefont {Young}},\ and\
  \bibinfo {author} {\bibfnamefont {R.}~\bibnamefont {Santra}},\ }\bibfield
  {title} {\bibinfo {title} {Impact of hollow-atom formation on coherent x-ray
  scattering at high intensity},\ }\href@noop {} {\bibfield  {journal}
  {\bibinfo  {journal} {Physical Review A}\ }\textbf {\bibinfo {volume} {83}},\
  \bibinfo {pages} {033402} (\bibinfo {year} {2011})}\BibitemShut {NoStop}%
\bibitem [{\citenamefont {Jurek}\ \emph {et~al.}(2016)\citenamefont {Jurek},
  \citenamefont {Son}, \citenamefont {Ziaja},\ and\ \citenamefont
  {Santra}}]{jurek2016xmdyn}%
  \BibitemOpen
  \bibfield  {author} {\bibinfo {author} {\bibfnamefont {Z.}~\bibnamefont
  {Jurek}}, \bibinfo {author} {\bibfnamefont {S.-K.}\ \bibnamefont {Son}},
  \bibinfo {author} {\bibfnamefont {B.}~\bibnamefont {Ziaja}},\ and\ \bibinfo
  {author} {\bibfnamefont {R.}~\bibnamefont {Santra}},\ }\bibfield  {title}
  {\bibinfo {title} {Xmdyn and xatom: versatile simulation tools for
  quantitative modeling of x-ray free-electron laser induced dynamics of
  matter},\ }\href@noop {} {\bibfield  {journal} {\bibinfo  {journal} {Journal
  of Applied Crystallography}\ }\textbf {\bibinfo {volume} {49}},\ \bibinfo
  {pages} {1048} (\bibinfo {year} {2016})}\BibitemShut {NoStop}%
\bibitem [{\citenamefont {Rohringer}\ \emph {et~al.}(2006)\citenamefont
  {Rohringer}, \citenamefont {Gordon},\ and\ \citenamefont
  {Santra}}]{rohringer2006configuration}%
  \BibitemOpen
  \bibfield  {author} {\bibinfo {author} {\bibfnamefont {N.}~\bibnamefont
  {Rohringer}}, \bibinfo {author} {\bibfnamefont {A.}~\bibnamefont {Gordon}},\
  and\ \bibinfo {author} {\bibfnamefont {R.}~\bibnamefont {Santra}},\
  }\bibfield  {title} {\bibinfo {title} {Configuration-interaction-based
  time-dependent orbital approach for ab initio treatment of electronic
  dynamics in a strong optical laser field},\ }\href@noop {} {\bibfield
  {journal} {\bibinfo  {journal} {Physical Review A}\ }\textbf {\bibinfo
  {volume} {74}},\ \bibinfo {pages} {043420} (\bibinfo {year}
  {2006})}\BibitemShut {NoStop}%
\bibitem [{\citenamefont {Rothman}\ \emph
  {et~al.}(2005{\natexlab{a}})\citenamefont {Rothman}, \citenamefont {Ho},\
  and\ \citenamefont {Rabitz}}]{rothman2005observable}%
  \BibitemOpen
  \bibfield  {author} {\bibinfo {author} {\bibfnamefont {A.}~\bibnamefont
  {Rothman}}, \bibinfo {author} {\bibfnamefont {T.-S.}\ \bibnamefont {Ho}},\
  and\ \bibinfo {author} {\bibfnamefont {H.}~\bibnamefont {Rabitz}},\
  }\bibfield  {title} {\bibinfo {title} {Observable-preserving control of
  quantum dynamics over a family of related systems},\ }\href@noop {}
  {\bibfield  {journal} {\bibinfo  {journal} {Physical Review A}\ }\textbf
  {\bibinfo {volume} {72}},\ \bibinfo {pages} {023416} (\bibinfo {year}
  {2005}{\natexlab{a}})}\BibitemShut {NoStop}%
\bibitem [{\citenamefont {Rothman}\ \emph
  {et~al.}(2005{\natexlab{b}})\citenamefont {Rothman}, \citenamefont {Ho},\
  and\ \citenamefont {Rabitz}}]{rothman2005quantum}%
  \BibitemOpen
  \bibfield  {author} {\bibinfo {author} {\bibfnamefont {A.}~\bibnamefont
  {Rothman}}, \bibinfo {author} {\bibfnamefont {T.-S.}\ \bibnamefont {Ho}},\
  and\ \bibinfo {author} {\bibfnamefont {H.}~\bibnamefont {Rabitz}},\
  }\bibfield  {title} {\bibinfo {title} {Quantum observable homotopy tracking
  control},\ }\href@noop {} {\bibfield  {journal} {\bibinfo  {journal} {The
  Journal of chemical physics}\ }\textbf {\bibinfo {volume} {123}},\ \bibinfo
  {pages} {134104} (\bibinfo {year} {2005}{\natexlab{b}})}\BibitemShut
  {NoStop}%
\bibitem [{\citenamefont {Rothman}\ \emph {et~al.}(2006)\citenamefont
  {Rothman}, \citenamefont {Ho},\ and\ \citenamefont
  {Rabitz}}]{rothman2006exploring}%
  \BibitemOpen
  \bibfield  {author} {\bibinfo {author} {\bibfnamefont {A.}~\bibnamefont
  {Rothman}}, \bibinfo {author} {\bibfnamefont {T.-S.}\ \bibnamefont {Ho}},\
  and\ \bibinfo {author} {\bibfnamefont {H.}~\bibnamefont {Rabitz}},\
  }\bibfield  {title} {\bibinfo {title} {Exploring the level sets of quantum
  control landscapes},\ }\href@noop {} {\bibfield  {journal} {\bibinfo
  {journal} {Physical Review A}\ }\textbf {\bibinfo {volume} {73}},\ \bibinfo
  {pages} {053401} (\bibinfo {year} {2006})}\BibitemShut {NoStop}%
\bibitem [{\citenamefont {Beltrani}\ \emph {et~al.}(2009)\citenamefont
  {Beltrani}, \citenamefont {Ghosh},\ and\ \citenamefont
  {Rabitz}}]{beltrani2009exploring}%
  \BibitemOpen
  \bibfield  {author} {\bibinfo {author} {\bibfnamefont {V.}~\bibnamefont
  {Beltrani}}, \bibinfo {author} {\bibfnamefont {P.}~\bibnamefont {Ghosh}},\
  and\ \bibinfo {author} {\bibfnamefont {H.}~\bibnamefont {Rabitz}},\
  }\bibfield  {title} {\bibinfo {title} {Exploring the capabilities of quantum
  optimal dynamic discrimination},\ }\href@noop {} {\bibfield  {journal}
  {\bibinfo  {journal} {The Journal of chemical physics}\ }\textbf {\bibinfo
  {volume} {130}},\ \bibinfo {pages} {164112} (\bibinfo {year}
  {2009})}\BibitemShut {NoStop}%
\bibitem [{\citenamefont {Dominy}\ and\ \citenamefont
  {Rabitz}(2012)}]{dominy2012dynamic}%
  \BibitemOpen
  \bibfield  {author} {\bibinfo {author} {\bibfnamefont {J.}~\bibnamefont
  {Dominy}}\ and\ \bibinfo {author} {\bibfnamefont {H.}~\bibnamefont
  {Rabitz}},\ }\bibfield  {title} {\bibinfo {title} {Dynamic homotopy and
  landscape dynamical set topology in quantum control},\ }\href@noop {}
  {\bibfield  {journal} {\bibinfo  {journal} {Journal of mathematical physics}\
  }\textbf {\bibinfo {volume} {53}},\ \bibinfo {pages} {082201} (\bibinfo
  {year} {2012})}\BibitemShut {NoStop}%
\bibitem [{\citenamefont {Miao}\ \emph {et~al.}(2000)\citenamefont {Miao},
  \citenamefont {Kirz},\ and\ \citenamefont {Sayre}}]{miao2000oversampling}%
  \BibitemOpen
  \bibfield  {author} {\bibinfo {author} {\bibfnamefont {J.}~\bibnamefont
  {Miao}}, \bibinfo {author} {\bibfnamefont {J.}~\bibnamefont {Kirz}},\ and\
  \bibinfo {author} {\bibfnamefont {D.}~\bibnamefont {Sayre}},\ }\bibfield
  {title} {\bibinfo {title} {The oversampling phasing method},\ }\href@noop {}
  {\bibfield  {journal} {\bibinfo  {journal} {Acta Crystallographica Section D:
  Biological Crystallography}\ }\textbf {\bibinfo {volume} {56}},\ \bibinfo
  {pages} {1312} (\bibinfo {year} {2000})}\BibitemShut {NoStop}%
\bibitem [{\citenamefont {Miao}\ and\ \citenamefont
  {Sayre}(2000)}]{miao2000possible}%
  \BibitemOpen
  \bibfield  {author} {\bibinfo {author} {\bibfnamefont {J.}~\bibnamefont
  {Miao}}\ and\ \bibinfo {author} {\bibfnamefont {D.}~\bibnamefont {Sayre}},\
  }\bibfield  {title} {\bibinfo {title} {On possible extensions of x-ray
  crystallography through diffraction-pattern oversampling},\ }\href@noop {}
  {\bibfield  {journal} {\bibinfo  {journal} {Acta Crystallographica Section A:
  Foundations of Crystallography}\ }\textbf {\bibinfo {volume} {56}},\ \bibinfo
  {pages} {596} (\bibinfo {year} {2000})}\BibitemShut {NoStop}%
\bibitem [{\citenamefont {Ye}\ \emph {et~al.}(2019)\citenamefont {Ye},
  \citenamefont {Rouxel}, \citenamefont {Cho},\ and\ \citenamefont
  {Mukamel}}]{ye2019imaging}%
  \BibitemOpen
  \bibfield  {author} {\bibinfo {author} {\bibfnamefont {L.}~\bibnamefont
  {Ye}}, \bibinfo {author} {\bibfnamefont {J.~R.}\ \bibnamefont {Rouxel}},
  \bibinfo {author} {\bibfnamefont {D.}~\bibnamefont {Cho}},\ and\ \bibinfo
  {author} {\bibfnamefont {S.}~\bibnamefont {Mukamel}},\ }\bibfield  {title}
  {\bibinfo {title} {Imaging electron-density fluctuations by multidimensional
  x-ray photon-coincidence diffraction},\ }\href@noop {} {\bibfield  {journal}
  {\bibinfo  {journal} {Proceedings of the National Academy of Sciences}\
  }\textbf {\bibinfo {volume} {116}},\ \bibinfo {pages} {395} (\bibinfo {year}
  {2019})}\BibitemShut {NoStop}%
\bibitem [{\citenamefont {Hosseinizadeh}\ \emph {et~al.}(2021)\citenamefont
  {Hosseinizadeh}, \citenamefont {Breckwoldt}, \citenamefont {Fung},
  \citenamefont {Sepehr}, \citenamefont {Schmidt}, \citenamefont {Schwander},
  \citenamefont {Santra},\ and\ \citenamefont
  {Ourmazd}}]{hosseinizadeh2021few}%
  \BibitemOpen
  \bibfield  {author} {\bibinfo {author} {\bibfnamefont {A.}~\bibnamefont
  {Hosseinizadeh}}, \bibinfo {author} {\bibfnamefont {N.}~\bibnamefont
  {Breckwoldt}}, \bibinfo {author} {\bibfnamefont {R.}~\bibnamefont {Fung}},
  \bibinfo {author} {\bibfnamefont {R.}~\bibnamefont {Sepehr}}, \bibinfo
  {author} {\bibfnamefont {M.}~\bibnamefont {Schmidt}}, \bibinfo {author}
  {\bibfnamefont {P.}~\bibnamefont {Schwander}}, \bibinfo {author}
  {\bibfnamefont {R.}~\bibnamefont {Santra}},\ and\ \bibinfo {author}
  {\bibfnamefont {A.}~\bibnamefont {Ourmazd}},\ }\bibfield  {title} {\bibinfo
  {title} {Few-fs resolution of a photoactive protein traversing a conical
  intersection},\ }\href@noop {} {\bibfield  {journal} {\bibinfo  {journal}
  {Nature}\ }\textbf {\bibinfo {volume} {599}},\ \bibinfo {pages} {697}
  (\bibinfo {year} {2021})}\BibitemShut {NoStop}%
\bibitem [{\citenamefont {Moths}\ and\ \citenamefont
  {Ourmazd}(2011)}]{moths2011bayesian}%
  \BibitemOpen
  \bibfield  {author} {\bibinfo {author} {\bibfnamefont {B.}~\bibnamefont
  {Moths}}\ and\ \bibinfo {author} {\bibfnamefont {A.}~\bibnamefont
  {Ourmazd}},\ }\bibfield  {title} {\bibinfo {title} {Bayesian algorithms for
  recovering structure from single-particle diffraction snapshots of unknown
  orientation: a comparison},\ }\href@noop {} {\bibfield  {journal} {\bibinfo
  {journal} {Acta Crystallographica Section A: Foundations of Crystallography}\
  }\textbf {\bibinfo {volume} {67}},\ \bibinfo {pages} {481} (\bibinfo {year}
  {2011})}\BibitemShut {NoStop}%
\bibitem [{\citenamefont {Cruz-Ch{\'u}}\ \emph {et~al.}(2021)\citenamefont
  {Cruz-Ch{\'u}}, \citenamefont {Hosseinizadeh}, \citenamefont {Mashayekhi},
  \citenamefont {Fung}, \citenamefont {Ourmazd},\ and\ \citenamefont
  {Schwander}}]{cruz2021selecting}%
  \BibitemOpen
  \bibfield  {author} {\bibinfo {author} {\bibfnamefont {E.~R.}\ \bibnamefont
  {Cruz-Ch{\'u}}}, \bibinfo {author} {\bibfnamefont {A.}~\bibnamefont
  {Hosseinizadeh}}, \bibinfo {author} {\bibfnamefont {G.}~\bibnamefont
  {Mashayekhi}}, \bibinfo {author} {\bibfnamefont {R.}~\bibnamefont {Fung}},
  \bibinfo {author} {\bibfnamefont {A.}~\bibnamefont {Ourmazd}},\ and\ \bibinfo
  {author} {\bibfnamefont {P.}~\bibnamefont {Schwander}},\ }\bibfield  {title}
  {\bibinfo {title} {Selecting xfel single-particle snapshots by geometric
  machine learning},\ }\href@noop {} {\bibfield  {journal} {\bibinfo  {journal}
  {Structural Dynamics}\ }\textbf {\bibinfo {volume} {8}},\ \bibinfo {pages}
  {014701} (\bibinfo {year} {2021})}\BibitemShut {NoStop}%
\bibitem [{\citenamefont {Fancher}\ \emph {et~al.}(2016)\citenamefont
  {Fancher}, \citenamefont {Han}, \citenamefont {Levin}, \citenamefont {Page},
  \citenamefont {Reich}, \citenamefont {Smith}, \citenamefont {Wilson},\ and\
  \citenamefont {Jones}}]{fancher2016use}%
  \BibitemOpen
  \bibfield  {author} {\bibinfo {author} {\bibfnamefont {C.~M.}\ \bibnamefont
  {Fancher}}, \bibinfo {author} {\bibfnamefont {Z.}~\bibnamefont {Han}},
  \bibinfo {author} {\bibfnamefont {I.}~\bibnamefont {Levin}}, \bibinfo
  {author} {\bibfnamefont {K.}~\bibnamefont {Page}}, \bibinfo {author}
  {\bibfnamefont {B.~J.}\ \bibnamefont {Reich}}, \bibinfo {author}
  {\bibfnamefont {R.~C.}\ \bibnamefont {Smith}}, \bibinfo {author}
  {\bibfnamefont {A.~G.}\ \bibnamefont {Wilson}},\ and\ \bibinfo {author}
  {\bibfnamefont {J.~L.}\ \bibnamefont {Jones}},\ }\bibfield  {title} {\bibinfo
  {title} {Use of bayesian inference in crystallographic structure refinement
  via full diffraction profile analysis},\ }\href@noop {} {\bibfield  {journal}
  {\bibinfo  {journal} {Scientific Reports}\ }\textbf {\bibinfo {volume} {6}},\
  \bibinfo {pages} {1} (\bibinfo {year} {2016})}\BibitemShut {NoStop}%
\bibitem [{\citenamefont {Ourmazd}\ \emph {et~al.}(2022)\citenamefont
  {Ourmazd}, \citenamefont {Moffat},\ and\ \citenamefont
  {Lattman}}]{ourmazd2022structural}%
  \BibitemOpen
  \bibfield  {author} {\bibinfo {author} {\bibfnamefont {A.}~\bibnamefont
  {Ourmazd}}, \bibinfo {author} {\bibfnamefont {K.}~\bibnamefont {Moffat}},\
  and\ \bibinfo {author} {\bibfnamefont {E.~E.}\ \bibnamefont {Lattman}},\
  }\bibfield  {title} {\bibinfo {title} {Structural biology is solved—now
  what?},\ }\href@noop {} {\bibfield  {journal} {\bibinfo  {journal} {Nature
  methods}\ }\textbf {\bibinfo {volume} {19}},\ \bibinfo {pages} {24} (\bibinfo
  {year} {2022})}\BibitemShut {NoStop}%
\bibitem [{\citenamefont {Bobkov}\ \emph {et~al.}(2015)\citenamefont {Bobkov},
  \citenamefont {Teslyuk}, \citenamefont {Kurta}, \citenamefont {Gorobtsov},
  \citenamefont {Yefanov}, \citenamefont {Ilyin}, \citenamefont {Senin},\ and\
  \citenamefont {Vartanyants}}]{bobkov2015sorting}%
  \BibitemOpen
  \bibfield  {author} {\bibinfo {author} {\bibfnamefont {S.}~\bibnamefont
  {Bobkov}}, \bibinfo {author} {\bibfnamefont {A.}~\bibnamefont {Teslyuk}},
  \bibinfo {author} {\bibfnamefont {R.}~\bibnamefont {Kurta}}, \bibinfo
  {author} {\bibfnamefont {O.~Y.}\ \bibnamefont {Gorobtsov}}, \bibinfo {author}
  {\bibfnamefont {O.}~\bibnamefont {Yefanov}}, \bibinfo {author} {\bibfnamefont
  {V.}~\bibnamefont {Ilyin}}, \bibinfo {author} {\bibfnamefont
  {R.}~\bibnamefont {Senin}},\ and\ \bibinfo {author} {\bibfnamefont
  {I.}~\bibnamefont {Vartanyants}},\ }\bibfield  {title} {\bibinfo {title}
  {Sorting algorithms for single-particle imaging experiments at x-ray
  free-electron lasers},\ }\href@noop {} {\bibfield  {journal} {\bibinfo
  {journal} {Journal of synchrotron radiation}\ }\textbf {\bibinfo {volume}
  {22}},\ \bibinfo {pages} {1345} (\bibinfo {year} {2015})}\BibitemShut
  {NoStop}%
\bibitem [{\citenamefont {Vartanyants}\ \emph {et~al.}(2007)\citenamefont
  {Vartanyants}, \citenamefont {Robinson}, \citenamefont {McNulty},
  \citenamefont {David}, \citenamefont {Wochner},\ and\ \citenamefont
  {Tschentscher}}]{vartanyants2007coherent}%
  \BibitemOpen
  \bibfield  {author} {\bibinfo {author} {\bibfnamefont {I.}~\bibnamefont
  {Vartanyants}}, \bibinfo {author} {\bibfnamefont {I.}~\bibnamefont
  {Robinson}}, \bibinfo {author} {\bibfnamefont {I.}~\bibnamefont {McNulty}},
  \bibinfo {author} {\bibfnamefont {C.}~\bibnamefont {David}}, \bibinfo
  {author} {\bibfnamefont {P.}~\bibnamefont {Wochner}},\ and\ \bibinfo {author}
  {\bibfnamefont {T.}~\bibnamefont {Tschentscher}},\ }\bibfield  {title}
  {\bibinfo {title} {Coherent x-ray scattering and lensless imaging at the
  european xfel facility},\ }\href@noop {} {\bibfield  {journal} {\bibinfo
  {journal} {Journal of Synchrotron Radiation}\ }\textbf {\bibinfo {volume}
  {14}},\ \bibinfo {pages} {453} (\bibinfo {year} {2007})}\BibitemShut
  {NoStop}%
\end{thebibliography}%
\end{document}